\documentclass[pdflatex,sn-standardnature,]{sn-jnl}
\usepackage{siunitx}
\usepackage{nicefrac}
\usepackage{amsmath}
\usepackage{ulem}
\usepackage{xcolor}
\usepackage{lineno}
\usepackage{graphicx}% Include figure files
\usepackage{dcolumn}% Align table columns on decimal point
\usepackage{bm}% bold math
\usepackage{placeins}
\definecolor{myRed}{RGB}{218,31,40}
\definecolor{myGreen}{RGB}{21,139,30}

\begin{document}

%\preprint{APS/123-QED}

\title{Common microscopic origin of the phase transitions in Ta$_2$NiS$_5$ and the excitonic insulator candidate Ta$_2$NiSe$_5$} 
%\linenumbers

%%%%%%%%%% Nature style affiliations %%%%%%%%%%%%

\author*[1]{Lukas Windg\"atter}
 \email{lukas.windgaetter@mpsd.mpg.de}

\author[2]{Malte R\"osner}
% \email{m.roesner@science.ru.nl}

\author[6]{Giacomo Mazza}
% \email{giacomo.mazza@unige.ch}

\author[1]{Hannes H\"ubener}
% \email{hannes.huebener@mpsd.mpg.de}

\author[3,4,5,6]{Antoine Georges}
% \email{ageorges@flatironinstitute.org}

\author[5,7]{Andrew J. Millis}
% \email{amillis@flatironinstitute.org}

\author*[1]{Simone Latini}
 \email{simone.latini@mpsd.mpg.de}

\author*[1,5,8]{Angel Rubio}
 \email{angel.rubio@mpsd.mpg.de}

\affil[1]{Max Planck Institute for the Structure and Dynamics of Matter, Luruper Chaussee 149, 22761 Hamburg, Germany}

\affil[2]{ Radboud University, Institute for Molecules and Materials, Heijendaalseweg 135, 6525 AJ Nijmegen, Netherlands}

\affil[3]{Coll{\`e}ge de France, 11 place Marcelin Berthelot, 75005 Paris, France}

\affil[4]{ Center for Computational Quantum Physics, Flatiron Institute, New York, New York 10010, USA}

\affil[5]{ CPHT, CNRS, Ecole Polytechnique, IP Paris, F-91128 Palaiseau, France}

\affil[6]{Department of Quantum Matter Physics, University of Geneva, Quai Ernest-Ansermet 24, 1211 Geneva, Switzerland}

\affil[7]{Department of Physics, Columbia University, New York, New York 10027, USA}

\affil[8]{Nano-Bio Spectroscopy Group, Departamento de F\'isica de Materiales, Universidad del Pa\'is Vasco, 20018 San Sebastian, Spain}

%%%%%%%%%% Nature affiliations %%%%%%%%%%%%

%\date{\today}

\abstract{
The structural phase transition in Ta$_2$NiSe$_5$ has been envisioned as driven by the formation of an excitonic insulating phase. However, the role of structural and electronic instabilities on crystal symmetry breaking has yet to be disentangled. Meanwhile, the phase transition in its complementary material Ta$_2$NiS$_5$ does not show any experimental hints of an excitonic insulating phase. We present a microscopic investigation of the electronic and phononic effects involved in the structural phase transition in Ta$_2$NiSe$_5$ and Ta$_2$NiS$_5$ using extensive first-principles calculations. In both materials the crystal symmetries are broken by phonon instabilities, which in turn lead to changes in the electronic bandstructure also observed in experiment. 
A total energy landscape analysis shows no tendency towards a purely electronic instability and we find that a sizeable lattice distortion is needed to open a bandgap. We conclude that an excitonic instability is not needed to explain the phase transition in both Ta$_2$NiSe$_5$ and Ta$_2$NiS$_5$.}

\maketitle

\newpage

\section{\label{Introduction}Introduction}

The excitonic insulator phase has been theoretically proposed in the 1960s \cite{Kohn1967,Jerome1967,Mott1961, Keldysh1968,Halperin1968,Kozlov1965} and is predicted to appear in semiconductors (or semimetals) with excitonic binding energies larger than the bandgap (or bandoverlap) of their conventional groundstate. Under this condition, the groundstate becomes unstable against the spontaneous formation of bound electron hole pairs, i.e. excitons. Depending on the conventional phase being semiconducting or semimetallic, the transition is described by a Bose-Einstein condensation (BEC) or a Bardeen-Cooper-Schrieffer (BCS) like mechanism respectively \cite{Keldysh1968,Jerome1967, Halperin1968}.

A major challenge in the experimental detection of an excitonic insulator state in a crystal is that the excitonic transition is coupled to other degrees of freedom, such as lattice distortions, which make an unambiguous detection difficult.
This is the reason why several materials such as 1T-TiSe$_2$ \cite{Cercellier2007, Kogar2017} or TmSe$_{0.45}$Te$_{0.55}$ \cite{TmSeTe1} have not been unambiguously confirmed to host an excitonic insulating (EI) groundstate. These are indirect bandgap or semi-metal materials which exhibit finite momentum ordering and in the case of  1T-TiSe$_2$, for example, the existence of a charge density wave masks the possible presence of an excitonic insulating phase. \newline
Very recently also two-dimensional materials have shown promises to host an excitonic groundstate. Experimentally, strong evidence for the existence of excitonic insulating phases have been provided both for transition metal dichalcogenides (TMD) bilayers \cite{KinFaiMak} and  monolayers such as WTe$_2$ and MoS$_2$ \cite{MonolayerWTe2, Varsano2020}. %in high quality transport measurements. 
An extensive theoretical study has identified several candidate material combinations for hetero-bilayers that could host an excitonic insulator by analyzing the electronic properties of a wide range of materials \cite{MaterialSearch}. 
Further low dimensional excitonic insulator candidates are carbon nanotubes \cite{Varsano2017}, Sb nanoflakes  \cite{Li2019}, double bilayer graphene \cite{Li2017_2, Perali2013} and topological systems \cite{Hu2018} such as InAs/GaSb \cite{Du2017}. 
In the bulk phase, MoS$_2$ has recently revealed as another candidate material since its bandgap can be tuned via pressure, yielding excitonic binding energies larger than the quasi-particle gap, which would allow for a transition into a possibly excitonic insulating groundstate \cite{MoS2Pressure}. \newline 
We investigate bulk Ta$_2$NiSe$_5$ (TNSe), which is considered to be a very promising candidate to host an excitonic condensate \cite{Kaneko2013}, with several experimental evidence suggesting an EI groundstate: Angle resolved photo electron spectroscopy (ARPES) measurements have shown a characteristic band flattening near the $\Gamma$-point upon cooling below the critical temperature \cite{Wakisaka2009,Seki2014,Watson2020,tang2020}, a dome-shaped bandgap-temperature phase diagram, similar to the theoretically predicted one, has been found \cite{Lu2017} and the opening of a gap has been measured in scanning tunnel spectroscopy (STS)\cite{Lee2019}, optics \cite{Lu2017} and ARPES \cite{Baldini2020, Watson2020} experiments below the critical temperature \cite{Lee2019}. The bandgap in the low-temperature phase is reported to be between 160 meV in optics\cite{Lu2017} and 300 meV in STS measurements \cite{Lee2019}. 
The current literature is controversially reporting the high temperature phase of TNSe to be either a direct bandgap semi-conductor \cite{Kaneko2012,Kaneko2013} or a small bandoverlap semi-metal\cite{Watson2020}. It is however agreed that there exists a structural phase transition, and possibly a transition into an excitonic groundstate, at the critical temperature T$_c$=326K \cite{Lu2017, Nakano2018, DiSalvo1986, Sunshine1985}. The structural transition has been reported to be from a high-temperature orthorhombic to a low-temperature monoclinic phase \cite{Nakano2018, DiSalvo1986, Sunshine1985}. \newline
On the other hand Ta$_2$NiS$_5$ (TNS), a material of the same structure of TNSe with Sulfur replacing Selenium, does not exhibit the same excitonic signatures of TNSe, such as band flattening and bandgap opening. Indeed a structural transition at 120K \cite{Blumberg2021} has been reported without the formation of flat bands or a metal to semiconductor transition \cite{Mu2018}.\newline %Therefore no excitonic insulating groundstate has been proposed for TNS \cite{Mu2018}. \newline

Due to the uncertainty on whether TNSe has a semi-metallic or gapped high-temperature phase, from a theoretical perspective it is unclear which mechanism between BEC and BCS can be applied. In fact experimentally there is no unambiguous evidence that excitons are involved in the phase transition at all.
These facts indicate that, in relation to TNSe,  the concept of excitonic insulator can be source of confusion, as it has been generally employed to refer to a phase showing a  characteristic gap opening and band flattening~\cite{Wakisaka2009,Seki2014,tang2020, Lee2019} without direct observation of excitonic features and/or excitonic condensation.

A recent symmetry analysis of the possible electronic instabilities issuing from the electron-electron interactions has shown that the electronic phase transitions in the material should not be be ascribed to a condensation phenomenon but rather to the electronic lowering of the discrete lattice symmetry~\cite{Mazza2020}, which corresponds to the breaking of a continuous U(1) symmetry during the excitonic phase transition. This observation, together with other recent experimental and theoretical observations \cite{Watson2020,subedi2020}, suggest a prominent role of structural instabilities in the system. 
In this work, we  abandon the concept of excitonic condensation and show that the experimental features of TNSe and TNS mentioned above can be completely explained without explicitly taking the formation of excitons into account.
We demonstrate how the electronic and structural properties of the two crystals can be rationalised in terms of electronic and intrinsic lattice instabilities that drive the system from the orthorhombic to monoclinic phase. The occurrence of such a lattice driven instability agrees with a prior report of Alaska Subedi \cite{subedi2020}.

We present a comprehensive ab-initio study using Density Functional Theory (DFT) for the transition from the orthorhombic to the monoclinic phase beyond standard generalized gradient approximation (GGA) functionals. In section \ref{sec:relaxation} we start by investigating the structural stability of the system and show, that at low temperature a monoclinic groundstate is energetically favoured to the orthorhombic one, which suggests the existence of a lattice instability. A similar analysis applied to TNS highlights an equivalent structural transition mechanism from an orthorhombic to a C2/c symmetric monoclinic cell.  \newline
In section \ref{sec:Electronic structure} we then investigate the electronic bandstructure of both the high and low-temperature phase of TNSe and TNS using different exchange correlation functionals with increasing accuracy. For TNSe, we can establish the orthorhombic groundstate to be metallic and observe a sizeable gap opening as well as a band flattening with the phase transition. This shows that for this material the structural distortion is essential for the bandgap opening observed in optics and STS measurements and cannot be explained considering only the electronic degrees of freedom. In section \ref{sec:GW} we show, that a similar finding results from our many body perturbation theory based calculations carried out at the G$_0$W$_0$ level. Performing an equivalent DFT analysis for TNS shows that both the orthorhombic and monoclinic phases are gapped systems with parabolic electronic dispersions at $\Gamma$. This explains why a metal to semiconductor transition and band flattening, which have been characterised as a signature of an excitonic insulator in TNSe, has not been observed for TNS \cite{Mu2018}.
We then show, that electronic heating enhances the valence band flattening and investigate in the following section \ref{sec: instability} the origin of the instability observed in TNSe. We find no evidence for an electronically-driven spontaneous symmetry breaking in the TNSe charge density upon infinitesimal lattice distortions. In the symmetry broken lattice structure the total energy is, however, clearly lowered, which renders the electron-phonon interaction the main driving mechanism for the structural transition. In section \ref{sec:phonons} we discuss the phononic bandstructure and single out the phononic instabilities of the orthorhombic phase, the ones that could possibly drive the structural transition. A detailed discussion of the obtained phonon modes in relation with the most recent Raman measurements shows that we are able to reproduce all phonon peaks, except for the first two B$_{\text{2g}}$ modes, which exhibit a strong broadening, which hints to a strong coupling of these modes to other degrees of freedom. Finally in section \ref{sec:ep-coupling} we show that the electronic dispersion is strongly modulated by different phonon modes. In particular it is expected that thermal occupation of the phonon mode responsible for the structural transition leads to a further bandgap opening %\newline

\section{Results}
\subsection{\label{sec:relaxation}Structure Relaxation}

TNSe is a layered material bound by van-der Waals forces. It consists of parallel Tantalum and Nickel chains and is thus referred to as a quasi 1D-material (see Fig.~1) \cite{Sunshine1985}. Experimentally the high-temperature phase exhibits an orthorhombic structure which is distorted to a monoclinic structure in the low-temperature phase. The distortion, however, is subtle with a change in the $\beta$ angle (see Fig.~2b) of just 0.53$^\circ$ \cite{Sunshine1985} to 0.69$^\circ$ \cite{Nakano2018}. % \newline

 \begin{figure}[ht]
     \centering
     \includegraphics[scale=0.4]{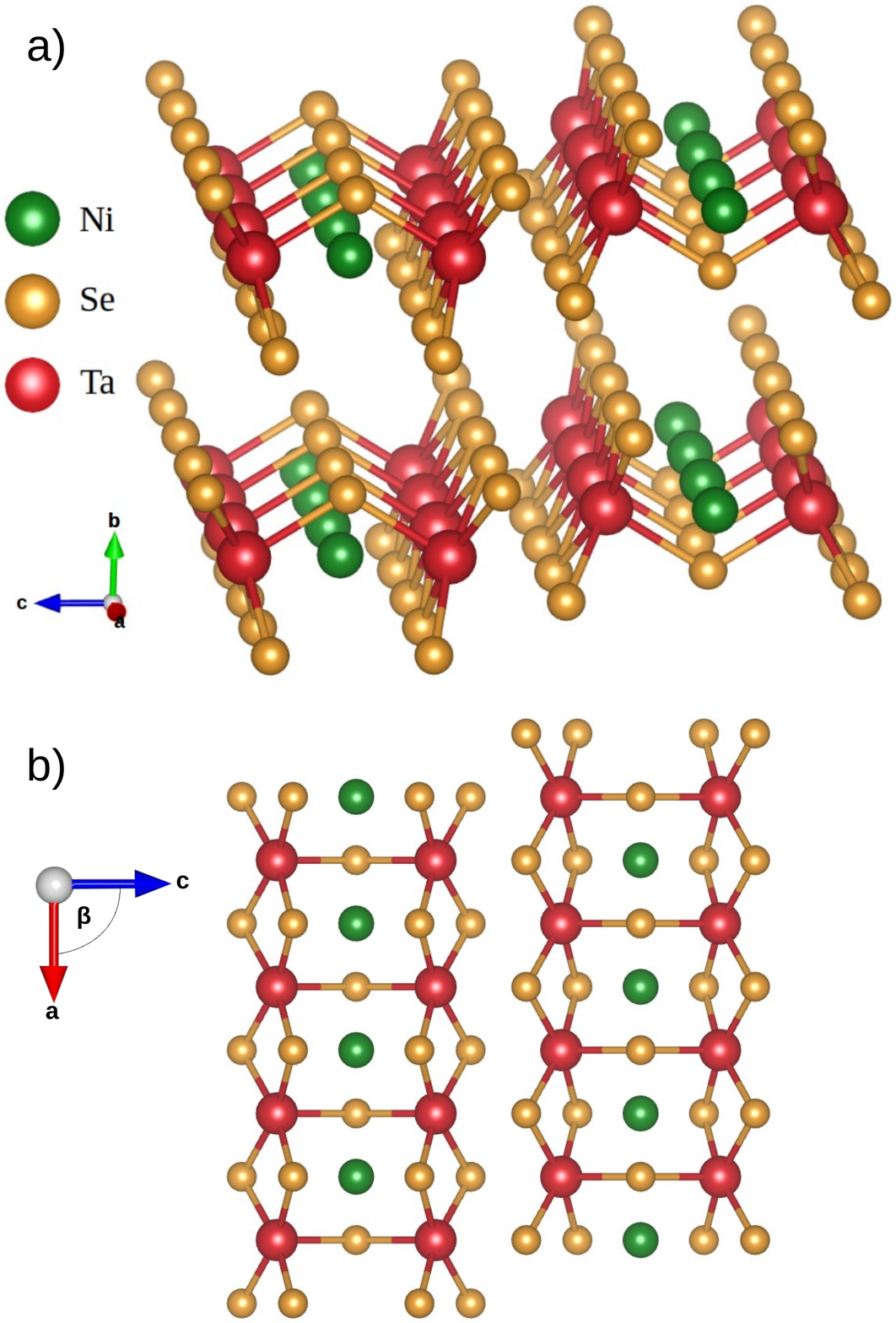}
     \caption{\textbf{Ta$_2$NiSe$_5$ Structure:} Layered structure of the orthorhombic phase of TNSe The Nickel and Tantalum atoms form parallel chains along the a-axis. Panel a) shows a 3D view of TNSe. Panel b) shows the projection onto the a-c-plane. For TNS the Se atoms are simply replaced with Sulfur atoms.
     }
     \label{fig:structure}
 \end{figure}

We have performed a first-principles structure relaxation of atomic coordinates, cell shape and cell volume for different functionals. In all cases the relaxation resulted in a triclinic structure. The resulting lattice parameters are shown in the Supplementary Tables 1 and 5: the two functionals vdW-optB88 and vdW-optPBE \cite{Klime2011,Klimes2010}, which include van-der Waals corrections, result in relaxed structures which agree well with the experimentally measured values, while the PBE functionals overestimates the experimentally measured interlayer lattice parameter b by 10.5\% \cite{Nakano2018} (see Supplementary Table 1 and 5). As expected this shows the importance of including van-der Waals forces when describing layered structures such as TNSe.

The angles of the relaxed triclinic structure are $\alpha = 90.005^{\circ}$, $\beta = 90.644^{\circ}$ and $\gamma = 89.948^{\circ}$. As the two angles $\alpha$ and $\gamma$ are almost rectangular, the triclinic cell and the corresponding monoclinic cell are almost degenerate. Furthermore, we have checked that the small triclinic distortion does not modify the electronic properties compared to the exact monoclinic structure. Thus in the following, we refer to the fully relaxed structure as monoclinic . The monoclinic angle $\beta=90.644^{\circ}$ agrees well with the experimentally measured values $90.693^{\circ}$ \cite{Nakano2018} and $90.53^\circ$\cite{Sunshine1985}. \newline

To obtain the relaxed geometry for the high-temperature orthorhombic cell, we have performed the same relaxation using the vdw-optB88 functional enforcing the orthorhombic symmetry. The resulting lattice parameters are shown in the Supplementary Tables 2 and 4. As the relaxed structure predicted by the vdW-optB88 functional has the best agreement with the experimentally measured values for both the monoclinic and orthorhombic phase, we have chosen it for all following structural calculations.

We have performed similar relaxations for the Sulfur compound: A full relaxation using the vdW-optB88 functional, again, results in a triclinic geometry that is almost degenerate with the monoclinic cell. The monoclinic angle is $\beta=90.50^\circ$. For the orthorhombic phase we have performed a further relaxation with fixed symmetry. The lattice parameters for both structures are shown in Supplementary Table 3 and agree very well with the experimentally measured values from X-ray diffraction.
This agrees with a recent Raman study by M.Ye et al which reports an orthorhombic high temperature phase and a structural transition to a monoclinic phase at T=120K \cite{Blumberg2021}.

Therefore we find, that in both TNSe and TNS a monoclinic geometry is energetically favoured, which shows that a similar lattice instability occurs in both materials. In section~\ref{sec:phonons} we extensively discuss the phononic properties of the two materials and identify the structural instability with an unstable B$_{2\text{g}}$ phonon.
Although the structural change between monoclinic and orthorhombic phase is small, the modifications for the electronic bandstructure with the phase transition is significant (see section~\ref{sec:Electronic structure}). This hints to a strong coupling between electronic and lattice degrees of freedom (see section~\ref{sec:ep-coupling}).

%%clearpage
\subsection{\label{sec:Electronic structure} Electronic Bandstructure}
In this section we present a complete and systematic DFT study of both TNSe and TNS using standard and more accurate functionals. While we find that the fine electronic properties are sensitive to the details of exchange and correlation, the underlying mechanism of bandgap opening due to the structural phase transition is independent of the functional choice for both TNSe and TNS. %\newline All this study is done at the theoretically calculated lattice constants described in section~\ref{sec:relaxation}. \newline

We  investigate the bandstructure of the orthorhombic and monoclinic geometries using different exchange correlation functionals which include varying degrees of exact exchange. These are ranked in chemical accuracy with the PBE functional \cite{Perdew1996} being the most inaccurate, but computationally least demanding, followed by the modified Becke-Johnson (mBJ) functional \cite{Tran2009,Becke2006,Becke1989} and finally the range separated hybrid functional HSE03 \cite{Heyd2003} being the most accurate one. 

The results are shown in Fig.~2.  Results for the HSE06 functional are reported Supplementary Figure 9 and the definition of the M-Z-$\Gamma$-X path is given in Supplementary Note 1. 
 \begin{figure*}[ht]
     \centering
     \includegraphics[scale=0.44]{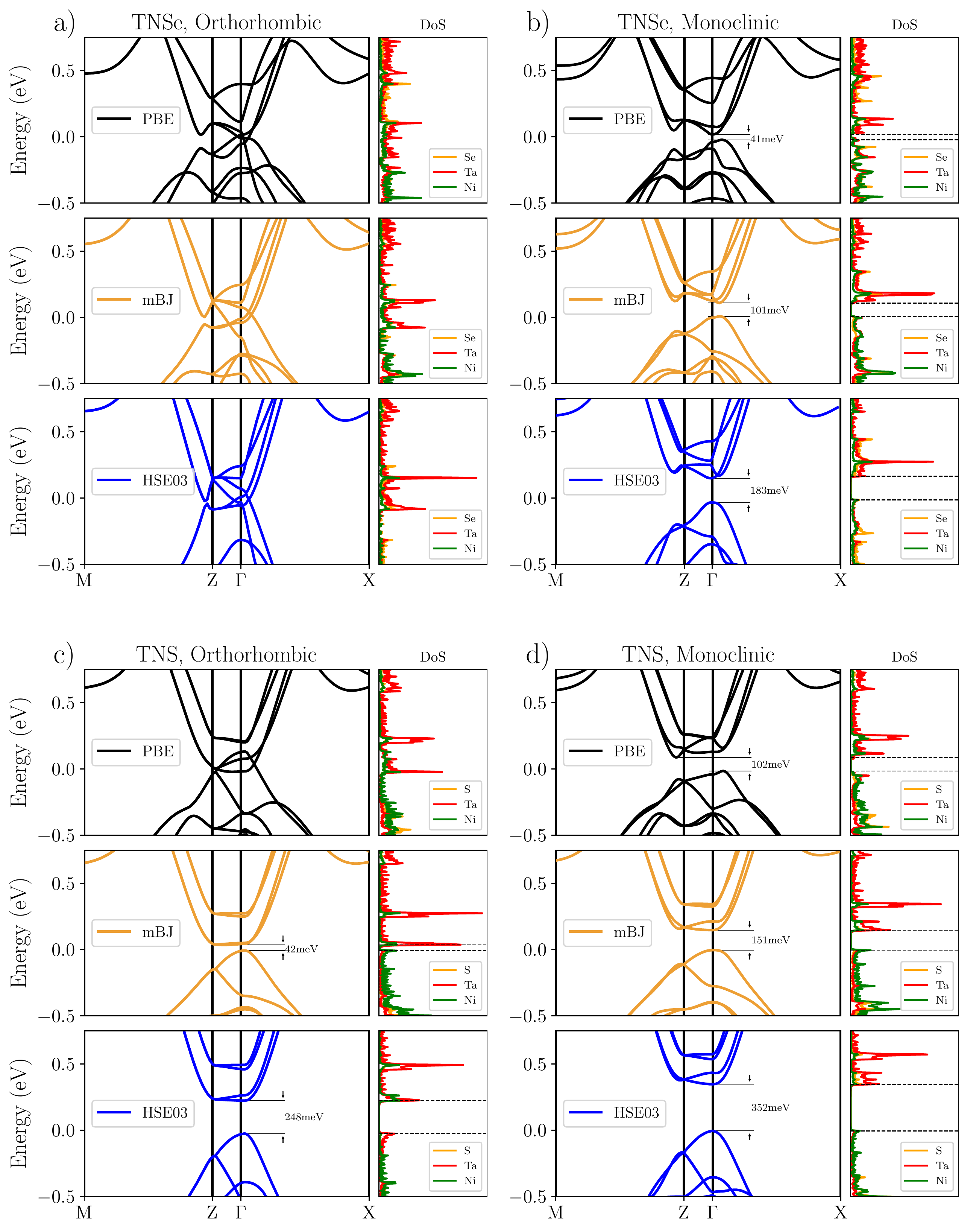}
     \caption{\textbf{Bandstructure of Ta$_2$NiSe$_5$ and Ta$_2$NiS$_5$:} Band structure of TNSe and TNS for the orthorhombic and monoclinic phase. For all investigated exchange correlation functionals we obtain a semi-metallic groundstate for the high temperature phase of TNSe. Only after the structural transition to the monoclinic geometry a bandgap of between 40 meV and 183 meV opens. In TNS already the orthorhombic phase exhibits a positive bandgap for the modified Becke Johnson and HSE03 functional which increases during the structural transition to the monoclinic phase. Therefore the metal to insulator transition observed in TNSe is lacking for TNS. In all plots, the Fermi-energy is set to 0 eV. The self consistently calculated c-Value for TNSe is c=1.25 and c=1.26 for TNS in both geometries. }
     \label{fig:dft-bands}
 \end{figure*}

%\subsection{TNSe}
Panels (a) and (b) of Fig.~2 display the obtained bandstructure of TNSe: the high temperature  orthorhombic structure exhibits a metallic groundstate independently of the functional presented here, while the monoclinic structure is always semi-conducting. 
The bandgap opening is sizeable for all investigated functionals and its position and value changes with 
the amount of exchange and correlation included. With the PBE functional we obtain an indirect bandgap value of 40 meV, 
with the modified Becke-Johnson (mBJ) functional a direct bandgap of 101 meV and with the HSE03 functional a direct bandgap of 183 meV. 
Thus, the bandgap grows with an increasing accuracy of the exchange and correlation functional. 
From Fig.~2b) we observe that in the monoclinic phase, 
TNSe has a rather flat dispersion around $\Gamma$ due to the valence band maxima being off $\Gamma$ for both the PBE and mBJ functional.

By increasing  the exchange contribution included in the functional, e.g. by tuning the mixing parameter in the hybrid functional or the c parameter in the mBJ-functional, it is possible to further increase the bandgap. We observed that even in the orthorhombic phase increasing 
the exchange contribution can lead to a gap opening (see Supplementary Note 4 and Supplementary Figures 5 and 6) which is consistent with prior studies \cite{Sugimoto2018}.
The bandgap opening resulting from HSE03 agrees well with the experimentally measured 
bandgap of 160 meV~\cite{Lu2017}. In STS measurements a gap of 300 meV has been reported~\cite{Lee2019}. 
However, we notice that the structure in Ref.\cite{Lee2019} shows a monoclinic distortion with an angle $\beta \propto$ 92.5$^\circ$,i.e. much larger than the angle reported in previous studies and the angle obtained in our calculation ($\beta \simeq 90.6^\circ$).
Correcting for this difference we find that both the optics and STM results are in good agreement with our HSE03 results (see Supplementary Note 6 for more details).

%\subsection{TNS}
Performing similar calculations for TNS, Fig.~2(c) and (d), we identify a major difference in the electronic behaviour of the Selenium and Sulfur compounds. 
We see that using either mBJ or HSE03 hybrid functionals, the calculations predict a semiconducting groundstate even for the orthorhombic geometry. The structural transition to the monoclinic cell then leads to a further bandgap opening, with bandgap values being around twice the size of the TNSe ones. Therefore, in contrast to TNSe, no metal to insulator transition is observed. The bandgap is direct at the $\Gamma$-point with a parabolic dispersion towards the X direction. 
These findings agree well with the theoretical and experimental reports: ARPES as well as optical conductivity data show that TNS has a parabolic dispersion around the $\Gamma$-point \mbox{\cite{Mu2018, Sugimoto2018}} with a robust bandgap of 200 meV \mbox{\cite{Mu2018,Li2017, Sugimoto2018}} and 250 meV in optics \mbox{\cite{Larkin2017}}. 

We stress that while both compounds have a similar lattice instability from orthorhombic to monoclinic geometry, in the case of TNSe the structural instability is accompanied by a metal-to-insulator transition which is absent in the case of TNS.
These observations match all known experimental  trends for these two systems.
In the case of the Selenium compound the metal-insulator transition accompanied by the flattening of the valence band around $\Gamma$ appears to be the intrinsic outcome of the structural instability of the system and not the distinctive signature of the excitonic insulator phase, as it has been interpreted so far. We support this statement in the next sections by explicitly accounting for the role of the electronic instability and comparing the excitonic binding energy with the bandgap value.
\newline
Given that states near the band gap have significant contributions from Ta, we have checked the validity of our results by also including spin-orbit coupling (SOC) effects and reported the related band structure calculations in the Supplementary Figures 2-4, Supplementary Note 3 and Supplementary Table 6. We find that besides a band splitting in the M-Z direction and a minor renormalization of the bandgap, all our claims remain unchanged.

%clearpage
\subsection{\label{sec:GW} G$_0$W$_0$ calculations}
While DFT calculations generally provide a good starting point for discussing material properties they often lack quantitative agreement with experimental results. It is known that one needs to add many body corrections to obtain a quantitative agreement with the experimental bandgaps. The standard approach which provides accurate bandgaps is the GW method  \cite{Schilfgaarde2006, Onida2002, Hybertsen1985, Hybertsen1986}. 
The latter is a many-body theory approach which consists in solving Hedin's equations iteratively \cite{Hedin1965} while neglecting vertex corrections. There exist various flavours of the GW approximation with the G$_0$W$_0$ being the most commonly used. In G$_0$W$_0$ one approximates the starting wavefunctions to be the DFT ones and performs only one cycle of Hedin's equation to compute the self energy and correct the electronic states energies. As such the G$_0$W$_0$ approximation relies on the DFT wavefunctions to provide a good description of the full many body wavefunction and it is crucial to investigate the starting point dependence of the G$_0$W$_0$ approximation. We have performed calculations using the wavefunctions and electronic dispersion resulting from PBE and HSE03 functional calculations (more hybrid functional starting points are shown Supplementary Figure 10). 
The corresponding bandstructures for the monoclinic geometry of TNSe can be seen in Fig.~3.

 \begin{figure}[t]
     \centering
     \includegraphics[scale=0.4]{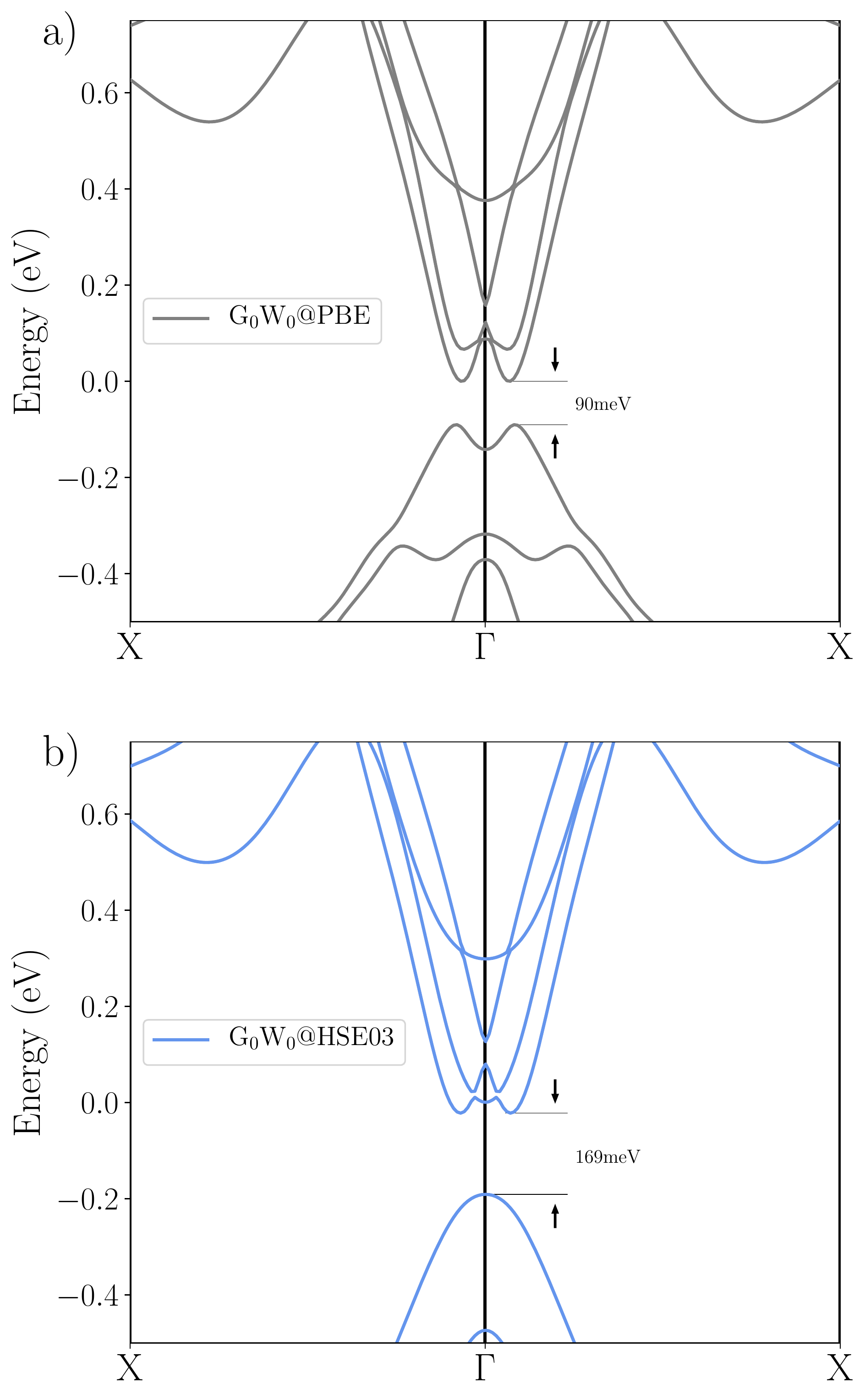}
     \caption{\textbf{G$_0$W$_0$ Bandstructure of monoclinic Ta$_2$NiSe$_5$:} Bandstructure of the monoclinic geometry of TNSe after performing a G$_0$W$_0$ calculation starting from the DFT groundstate of TNSe using the PBE (panel a) and the HSE03 hybrid functional (panel b). While the bandgaps are all within a small range of 100 meV to 163 meV, the valence band dispersions differ depending on the functional used for the initial wavefunctions. Panel a) has been adapted from our work in ref.\cite{Baldini2020}
     }
     \label{fig: startingPoint}
 \end{figure}

Independently of the starting point, the bandgap converges towards a similar value, increasing in the PBE case, where it is commonly underestimated, and decreasing for the HSE03 functional. In all cases G$_0$W$_0$ predicts a bandgap between 100 meV and 163 meV which is in good agreement with the experimental gap of 160 meV measured in optics \cite{Lu2017}. The dispersion of the conduction bands shows similar features with a pronounced double well shape in X-$\Gamma$-X direction. The valence band dispersion, however, differs between the PBE and the HSE, with the PBE one being M-shaped and the HSE one being parabolic. This is inherited from the different initial DFT wavefunctions and electronic dispersion. 

The comparison of the electronic dispersion with recent ARPES measurements by Baldini et al.\cite{Baldini2020} shows that the dispersion obtained using the PBE functional as starting point agrees well with the experimentally measured one, reproducing both the flat valence band as well as the Mexican hat shaped conduction band. Also Tang et al. \cite{tang2020} and Watson et al. \cite{Watson2020} report an M-shaped valence band which agrees well with the G$_0$W$_0$@PBE calculation. This reinforces the conclusion that the band structure in the monoclinic phase correctly 
accounts for the metal-to-insulator transition. A detailed quantitative comparison to experimental 
results, however, can only be achieved by including correlation beyond the PBE functional.
\newline

We also solved the Bethe-Salpether equation (BSE) starting from the G$_0$W$_0$ results \cite{Salpeter1951,Sander2015} to explicitly account for exciton formation. While full convergence is not feasible with our computational resources, we can extrapolate from our k-point dependent results BSE results on top of G$_0$W$_0$ and verify that the excitonic binding energy is smaller than the G$_0$W$_0$ gap in the monoclinic phase (see Supplementary Notes 7 and 8 and Supplementary Figures 13-19). This is an indication that the condition needed for an excitonic insulating phase, i.e. an exciton binding energy larger than the gap, is not fulfilled in TNSe for the monoclinic phase. 

%clearpage
\subsection{\label{sec:heating}The effect of electronic Heating}
The interpretation of an excitonic insulator as a phase consisting of an excitonic condensate entails that an increase of temperature above a critical value leads to the melting of the condensate and hence to a bandgap closing. Experiments aimed at demonstrating such a behaviour to prove the possibility of excitonic condensation in TNSe, have indeed reported bandgap opening and band flattening for increasing pump fluence \cite{tang2020}. Here we mimic the photoexcitation with a thermal distribution of electrons and holes and show that the variation of the bandgap and the band flattening can be explained via a temperature increase in the same manner as in standard semiconductors. Indeed, an increase of temperature of the thermal distribution generates free carriers and it has been shown that for low carrier densities the bandgap shrinks and eventually increases again for sufficiently high densities \cite{Sparatu2004,Beni1978}. The mechanism behind such a bandgap variation can be rationalized in terms of effective screening of the Coulomb interaction which modulates the effect of bandgap opening caused by the exchange contribution.

 \begin{figure*}
     \centering
     \includegraphics[scale=0.28]{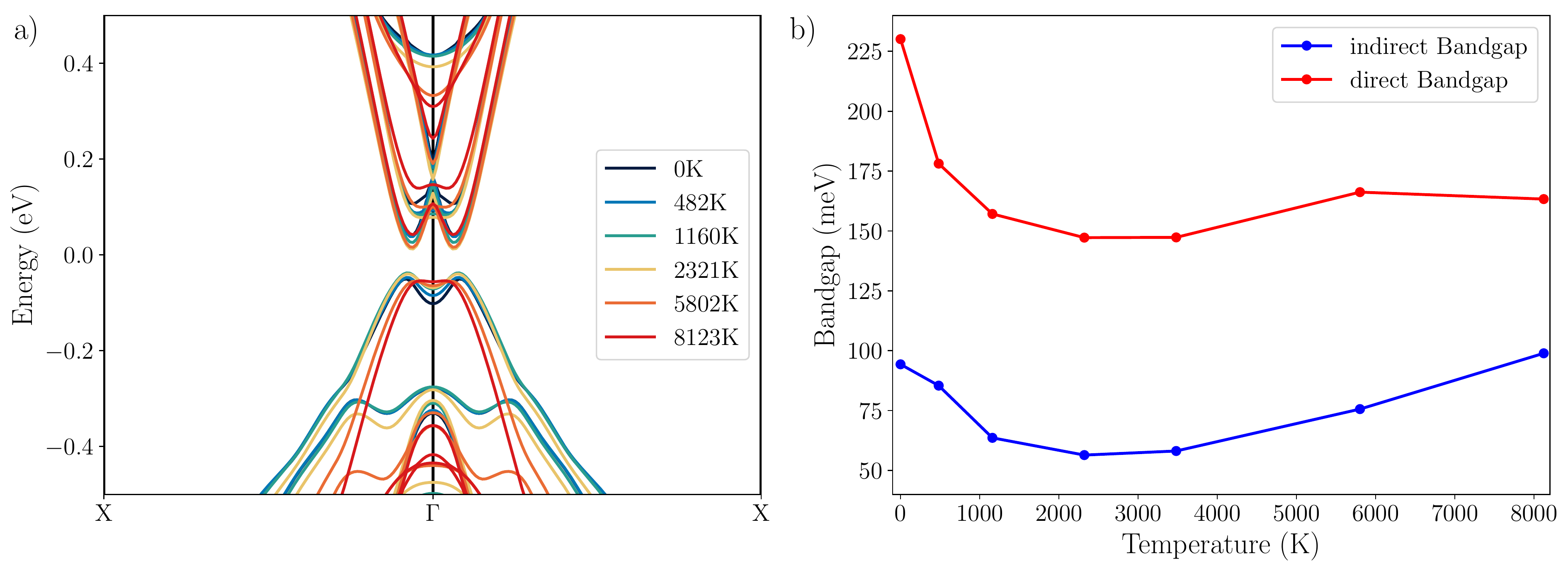}
     \caption{\textbf{Bandstructure of Ta$_2$NiSe$_5$ for different electronic temperatures:} DFT+G$_0$W$_0$ calculation using different electronic smearing temperatures. a) Shows the electronic dispersion with increased temperature. The obtained bandstructures agree very well with the flat bands obtained in ARPES measurements. b) Shows the evolution of the bandgap with increasing electronic temperature. It decreases rapidly to ~50\% of its equilibrium value at 2000K and increases afterwards due to an increased band flattening for very high temperatures. The direct bandgap has been measured at $\Gamma$. The figure has been adapted from our work in ref.\cite{Baldini2020}.}
     \label{fig: T-Smearing}
 \end{figure*}

To approximately investigate the enhanced screening effects due to free carriers, we have performed a set of G$_0$W$_0$ calculations for different electronic temperatures. The electronic temperatures have been set by changing the temperature in the Fermi-Dirac function used for the occupation smearing of the Kohn-Sham states in the self-consistent DFT calculations. This effectively leads to a depletion of electrons in the valence bands and a finite electronic occupancy of the conduction bands. Specifically we have investigated Fermi-Dirac distributions with temperatures ranging between $0$~K and $8000$~K. G$_0$W$_0$ calculations on such thermalized groundstates can then provide bandstructures which include the effect of the screening due to the thermally excited carriers. The resulting bandstructures and bandgap renormalizations are presented in Fig.~4(a) and Fig.~4(b). One observes that introducing charged carriers to the conduction bands rapidly decreases the bandgap. It reaches its minimum around 2500K with a bandgap $50\%$ smaller than its equilibrium value and reopens for higher temperatures. This behaviour is consistent with the typical bandgap renormalization in standard semiconductors mentioned above. The pronounced double well feature of the valence band slightly flattens out as the $\Gamma$-point energy is shifted upwards (see Fig.~4(a)) and only for electronic temperature higher than $4000$~K the bandstructure changes significantly, which explains the upwards trend in the bandgap for high temperatures. To reach such high electronic temperatures with a laser, however, is experimentally unrealistic as it would damage the material.

We remark that, even though the applied computational procedure does not properly describe the experimental photoexcitation of electrons into conduction bands, we could expect a behaviour similar to the one described above upon intensive laser pumping.

\subsection{\label{sec: instability}Origin of the Structural Instability}
Having established the importance of the structural instability in the determination of the insulating ground state of TNSe, in this section we discuss  possible origins of the structural instability. 
On general grounds, the structural instability is the result of the complex interplay between the electronic and the lattice degrees of freedom.
The possibility of a purely electronic origin of the structural  instability, namely a charge density that spontaneously  breaks the  orthorhombic symmetry at fixed ions position, has been  recently  discussed in the literature.
Specifically, this possibility has been introduced at the model level in  Ref.~\cite{Mazza2020} as the result of an electronic interaction-driven  spontaneous hybridization between localized Wannier orbitals of different symmetry.
Moreover, experimental data from symmetry-resolved Raman have  been interpreted as signatures of quasi-critical charge fluctuations that are active in the corresponding symmetry channel~\cite{Kim2020}.
 
We start by recalling the mechanism behind such an electronic instability. The mechanism can be readily understood in terms of the energetic competition between on-site (Ta-Ta and Ni-Ni) and the nearest neighbouring (Ta-Ni) density-density interaction that can render the ground state of the system to be either metallic or insulating.
In a simplified description this energetic competition can be well understood as a problem of two electrons in two sites with on-site interaction $U$ and inter-site interaction $V$.
If the energy of the lowest lying orbitals in the two sites are comparable, the ground state configuration of the problem is entirely determined by the ratio between $U$ and $V$.
In the limit $U \gg V$ the lowest energy configuration corresponds to the single occupation of the lowest lying orbitals on the two sites, whereas in the opposite limit $U \ll V$ the most favourable configuration is the doubly occupation of the site with the lowest energy orbital.
The two configurations become quasi-degenerate for $U \sim V$ such that a coherent superposition between the two configuration due to an exchange interaction mediated by $V$ is possible.

Assuming that in the extended system the lowest lying orbitals in the two sites contribute to the formation of two bands close to the Fermi level, the two configurations correspond respectively 
to a metal in which  both bands are partially occupied or to an insulator in which the lowest lying band is fully occupied. The coherent superposition, instead, indicates the possibility of a spontaneous hybridization between the bands which leads to the opening of an hybridization gap.
As discussed in Ref.~\cite{Mazza2020} this can happen only if the bands have different symmetry character and therefore an hybridization of this type would correspond to a charge density that breaks the orthorhombic lattice symmetry.
Calculations based on a mean-field ansatz for a realistic model of TNSe show that the spontaneous hybridization transition can occur in a very narrow region around $U \sim 4 V$ (see inset of Fig.~5), where the factor $4$ comes from the fact  that there are four neighbouring Ta-atoms around each Nickel. We emphasize that the phase diagram in the inset of Fig.~5) from Ref.~\cite{Mazza2020} is intended to be qualitative as it is specifically calculated for a set of realistic parameters corresponding to the mBJ functional and a selection of Wannier orbitals, hence  not necessarily representative for other functionals.  \newline
The smallness of this region suggests that the electronic instability relies on a delicate energetic balance between the competing metallic and insulating states which is not guaranteed to be exactly satisfied in the real material.
We therefore exploit the present full ab-initio results to test this energetic balance in the case of TNSe and, in particular, how it is affected by the lattice distortions which are neglected in the electronic instability discussion above.
We do so by performing bandgap calculations, reported in Fig.~5, for different functionals on increasingly distorted structures using the distortion parameter $d$. To define the distortion parameter we linearly interpolate between the high symmetric orthorhombic and low symmetric monoclinic geometry: 
we parameterized this transition with a transition vector $\textbf{v}_\text{t}$ defined as the difference between the atomic configuration of the high symmetry and low symmetry structure $\textbf{v}_\text{t} = \textbf{v}_{\text{m}} - \textbf{v}_{\text{o}}$, with $\textbf{v}_{\text{m}}$ describing the atomic configuration of the monoclinic cell and $\textbf{v}_{\text{o}}$ describing the atomic configuration of the orthorhombic cell. 
Using this vector we can linearly interpolate between the orthorhombic and monoclinic structure via $\textbf{v}_\text{i}(d) = \textbf{v}_\text{o} + d \cdot \textbf{v}_\text{t}$, where $d$ is the distortion parameter. The lattice vectors are parameterized in a similar way. This means, that also each of the three lattice vectors of both geometries are linearly interpolated using the difference between the monoclinic and orthorhombic lattice vectors 
$\textbf{v}_\text{lat,t} = \textbf{v}_\text{lat,m} - \textbf{v}_\text{lat,o}$. Hence, the interpolated lattice vectors are $\textbf{v}_\text{lat,i}(d) = \textbf{v}_\text{lat,o} + d \cdot \textbf{v}_\text{lat,t}$ for each of the three lattice vectors. 
If a gap opens as a result of the electronic instability only, we could expect that an infinitesimal distortion $d \to 0^+$ would be sufficient  to stabilize an insulating ground state with a distorted charge density. 
However, we observe that, independently of the functional used, a finite critical value of the distortion parameter $d_c$ is needed in order to open a gap (see Fig.~5).
This means that TNSe is indeed likely outside the region where a purely electronic instability can be expected and that a finite distortion of the lattice is critical to account for the metal-insulator transition.
Nonetheless, we observe a cooperation between electronic correlations and lattice distortion in the opening of the gap.
Indeed, by increasing the accuracy of the functional (from PBE to HSE), which corresponds to increasing amount of electronic exchange and correlation, 
the bandgap at fixed distortion is increased and the critical value of the distortion parameter $d_c$ needed for opening a gap is significantly reduced.

We therefore conclude that in relation to the simple mean-field phase diagram with tunable $U$ and $V$ interaction derived for the orthorhombic phase of TNSe, the material must be placed 
on the semi-metallic region of the phase diagram and that the lattice distortions act in the same direction of the nearest neighbour interaction $V$.
On the other hand, based on the previous electronic band structure results, the Sulfur compound must be placed on the semiconducting side of the  phase diagram, corresponding to a larger effect of the interaction $V$.

As a further confirmation of the fact that the non-local interaction $V$ is responsible for the opening of the gap in these materials we also performed fully ab-initio Hartree-Fock calculations for TNSe which are supposed to include a larger contribution from non-local  correlations: both the orthorhombic and the monoclinic phase have semiconducting groundstates with bandgaps of $436$~meV and $836$~meV respectively (see Supplementary Figure 11).

 \begin{figure}[t]
     \centering
     \includegraphics[scale=0.35]{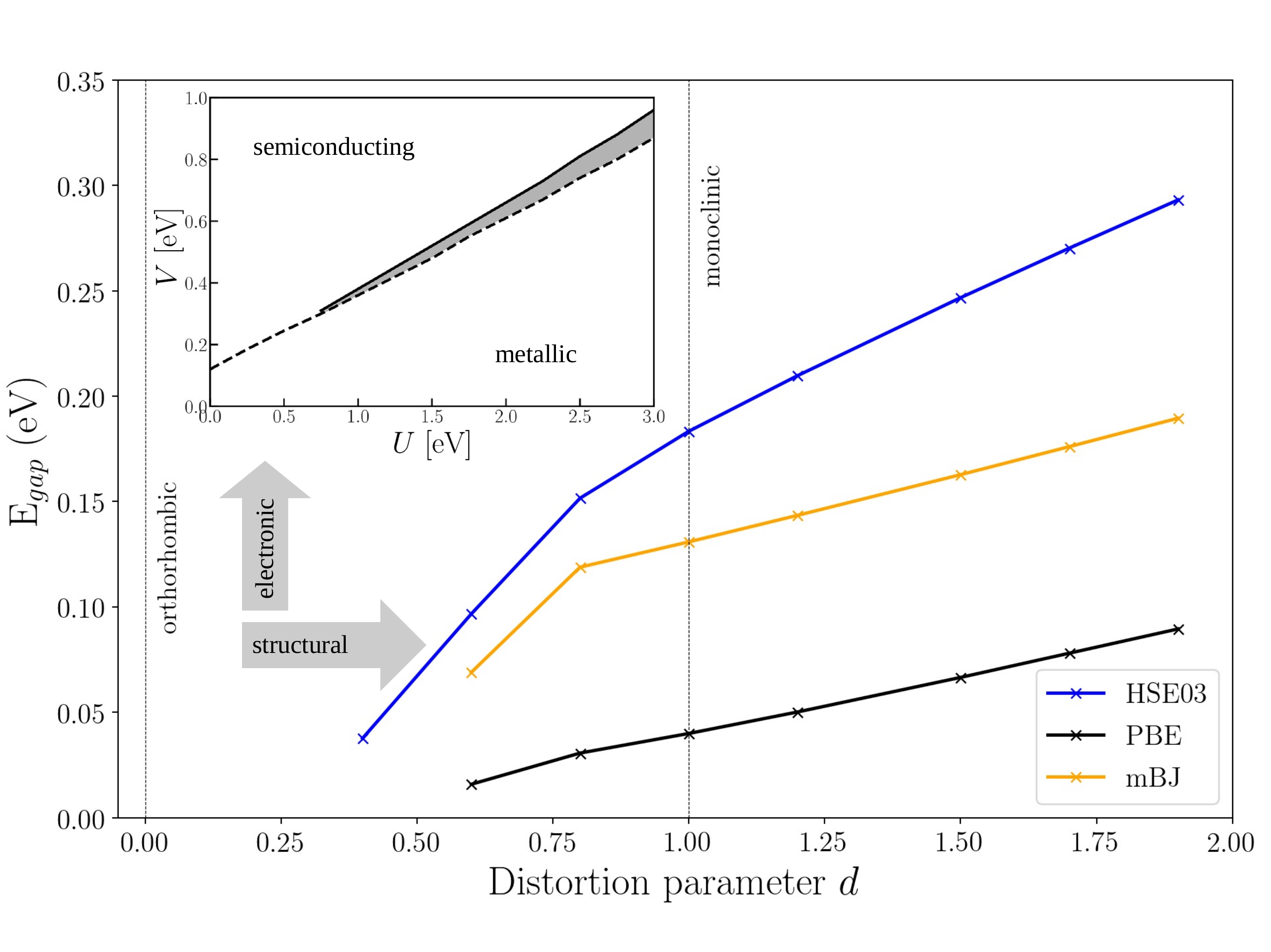}
     \caption{\textbf{Bandgap of Ta$_2$NiSe$_5$ along the phase transition:} Bandgap as a function of the distortion parameter $d$ defined in the text for different exchange correlation functionals for TNSe: for all considered exchange functionals we obtain a positive bandgap only for finite distortions $d$ which then increases as a function of the distortion parameter and the functional chosen. A minimal distortion breaking all relevant lattice symmetries ( $d=$0.05) does not open a bandgap which would be expected for an electronically driven instability (see Supplementary Figure 8). We displayed only data points which have a finite bandgap. Gray arrows indicates the bandgap opening due to structural distortion and electronic correlation using different functionals. The inset shows the Phase Diagram of TNSe adapted from \cite{Mazza2020}. The region between dashed and solid (gray shaded) line shows the region with an electronic instability breaking the orthorhombic lattice symmetry towards the monoclinic. Above is the semi-conducting region without electronic symmetry breaking and below the metallic region.
     }
     \label{fig:gapEvolution}
 \end{figure}

We finally mention that despite the fact that a purely electronic contribution is absent for TNSe we cannot exclude the system to be close enough to an electronic instability so that quasi-critical charge fluctuations can be present in the symmetry channel corresponding to the  breaking of the mirror symmetries  that characterizes the transition from the  orthorhombic to the monoclinic phase~\cite{Kim2020}. 
This analysis goes beyond the scope of the present work and will be the subject of future investigations.
\newline

In the following we focus on the possibility of having an energetically favorable electronic instability which however does not open a bandgap, but instead stays in the metallic regime close to the orthorhombic geometry. 
We introduce the total energy $E\equiv E \left[d,\rho_- \right]$, as a function of two coordinates. The first one, which accounts for the lattice degrees of freedom, is the lattice distortion $d$. 
The second coordinate of the energy functional is the 
component $\rho_{-}$ of the electronic charge density which corresponds to a symmetry lowering from orthorhombic to monoclinic (of \cite{Mazza2020}). It is defined as

\begin{equation}
    \rho_{-} = \int\displaylimits_{x \in \text{unit cell}} \vert\rho(x) - \rho(R(x))\vert \, \, \mathrm{d}x  \quad , %\quad \text{with}\\
\end{equation}

with R being the orthorhombic symmetry operators and $\rho$ the charge density.
For the orthorhombic cmcm geometry of TNSe four different mirror symmetries $R=\lbrace \sigma_{A,o}, \sigma_{A,p}, \sigma_{B,o}, \sigma_{B,p} \rbrace $ have been identified as relevant to the orthorhombic to monoclinic transition \cite{Mazza2020}, where and $p$ and $o$ denote mirror symmetries parallel and orthogonal to the chain direction and $A$ and $B$ denote the corresponding  one dimensional chains in the unit cell. We will focus on these mirror symmetries in the following. 
In this picture the groundstate is defined such that at $d^*$ and $\rho_{-}^*$ the total energy is minimal:

\begin{equation}
    \frac{\partial E}{\partial d}\Bigr\vert_{d=d^*} = \frac{\partial E}{\partial \rho_{-}}\Bigr\vert_{\rho_{-}=\rho_{-}^*} = 0.
\end{equation}

If the system undergoes a spontaneous electronic instability, the charge density breaks the orthorhombic lattice symmetry for infinitesimal distortions and hence:

\begin{equation}
    \lim_{d \to 0} \rho_-^*(d) \neq 0.
\end{equation}

The two possible scenarios for a lattice and an electronically driven transition are shown schematically in Fig.~6. Panel a) shows the case of a lattice driven phase transition. The charge density component $\rho_{-}$ does not break the orthorhombic symmetries at $d=$0 and evolves smoothly towards the groundstate of the system with increasing distortion. Panel b) shows the case of an electronically driven phase transition where the charge density breaks the orthorhombic lattice symmetries. In this case the charge density component $\rho_{-}$ does not vanish at $d=$0  but it is finite instead and its value increases with increasing distortion . In both cases a) and b) a finite d-distortion may still be needed to obtain a positive bandgap.    

When we perform ab-initio calculations imposing the orthorhombic symmetries, the charge density is automatically symmetrized so that $\rho_{-}^*(d=0)$ is true by construction. However, if we compute $\rho_{-}$ for small lattice distortions that break the orthorhombic symmetry, we can evaluate the limit $\lim_{d \to 0} \rho_{-}^*(d)$ and verify whether an electronic instability is present. Furthermore, an electronic symmetry breaking should also result in a discontinuity in the total energy for infinitesimal lattice distortion, as a phase with a finite $\rho_{-}^*$ has to lower the energy with respect to the symmetric orthorhombic phase in the electronic instability scenario. Both the computed total energy as well as $\rho_{-}$ for fixed lattice distortions $d$ are depicted in panels c) and d) of Fig.~6.

 \begin{figure*}[ht]
     \centering
     \includegraphics[scale=0.35]{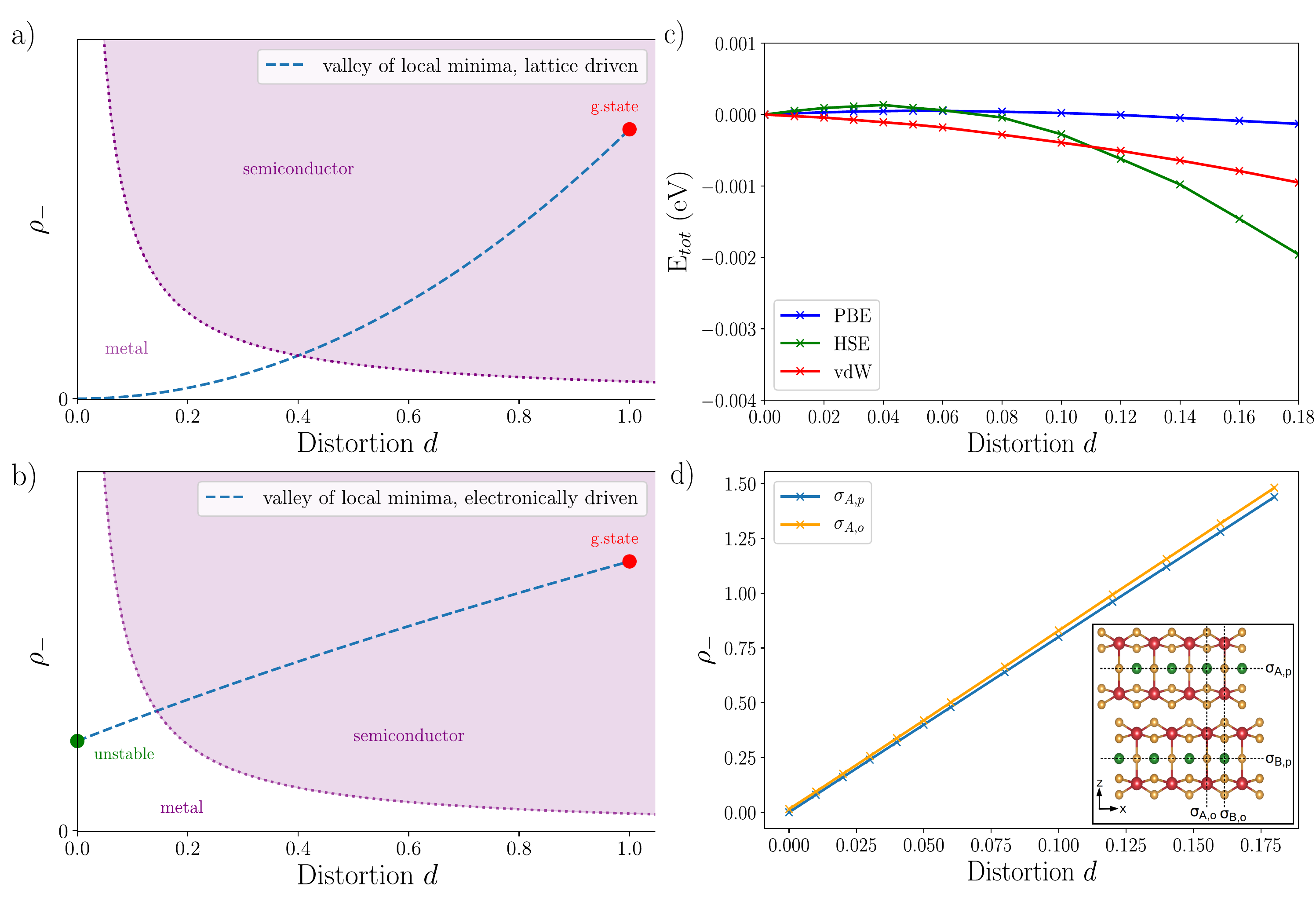}
     \caption{\textbf{Origin of the phase transition:} Sketch of the energy landscape of $E \left[d,\rho_- \right]$ along the local minima valleys for (a) a lattice driven and (b) an electronically driven phase transition. The green dot marks the groundstate defined via $d^*$ and $\rho_{-}^*$. In the case of a lattice driven phase transition (a) the charge density component $\rho_{-}$ evolves smoothly from 0 to $\rho_{-}^*$ as the lattice is distorted. In the case of an electronically driven phase transition the charge density breaks the orthorhombic symmetries for infinitesimal distortions. Thus, $\rho_{-}$ jumps discontinuously to a finite value (green dot marked as unstable) and evolves afterwards smoothly to the groundstate. The purple line marks, that even with an electronic instability a finite $d$ distortion may in general be needed to obtain a positive bandgap (c) Total energy as a function of the lattice distortion for the PBE, vdw-optB88 and HSE03 functional. The total energy of the respective orthorhombic geometry was taken as reference value and set to 0. (d) Increase of the symmetry breaking parameter $\rho_{-}$ with increasing lattice distortion using the HSE03 functional. If there was an electronic instability, both curves would have exhibited a discontinuity for infinitesimal lattice distortions d, with $\rho_{-}$ extrapolating to a positive non-zero value. A linear fit of both curves is given by $f_{\sigma_{\text{A,o}}}(d)=7.9964(\pm 0.0041) \cdot d+2.2\mathrm{e}{-4}(\pm 4.1\mathrm{e}{-4})$ and $f_{\sigma_{\text{A,p}}}(d)=8.1658(\pm 0.0120) \cdot d + 2.64\mathrm{e}{-3}(\pm 0.26\mathrm{e}{-3})$. The inset shows the orthorhombic symmetry planes defined by Mazza et al \cite{Mazza2020}.
     }
     \label{fig:tot_sym}
 \end{figure*}

Panel (c) shows, that there is no discontinuity in the total energy. For the vdw-optB88 functional the total energy decreases monotonously to its minimum for increasing $d$ (see Supplementary Note 2 and Supplementary Figure 1 for $q>0.2$). For the PBE and HSE functionals it increases first and subsequently decreases towards its minimum, which for the HSE03 functional is at around $d=1.6$. We attribute such a non-monotonous behavior to the fact that the distortion parameter $d$ is defined with respect to the vdw-optB88 functional structures. This means that in the phase-space of the normal phononic mode we are not moving along the soft phonon modes but along a direction where the orthorhombic phase is a saddle point. Solving this issue would require a structure relaxation using hybrid functionals to obtain the exact soft mode coordinate, which is computationally not feasible for TNSe.  Panel (d) shows the evolution of the charge density component $\rho_{-}$ for different symmetry operations defined in Ref.~\cite{Mazza2020}. $\sigma_{\text{A,p}}$ describes the reflection symmetry parallel the Nickel chain in the xz-plane and $\sigma_{\text{A,o}}$ describes the reflection in the yz-plane orthogonal to the chain direction and along the Tantalum atoms (see inset Fig.~6). For both reflection symmetries $\rho_-$ approaches 0 for $d$ $\rightarrow 0$ . 
%as the system is tend towards the higher symmetry orthorhombic phase.
This reinforces the conclusion that the transition in TNSe requires a finite lattice distortion and we do not observe a spontaneous electronic instability.

\subsection{Phonon Dispersion}
\label{sec:phonons}

 \begin{figure*}[t]
     \centering
     \includegraphics[scale=0.55]{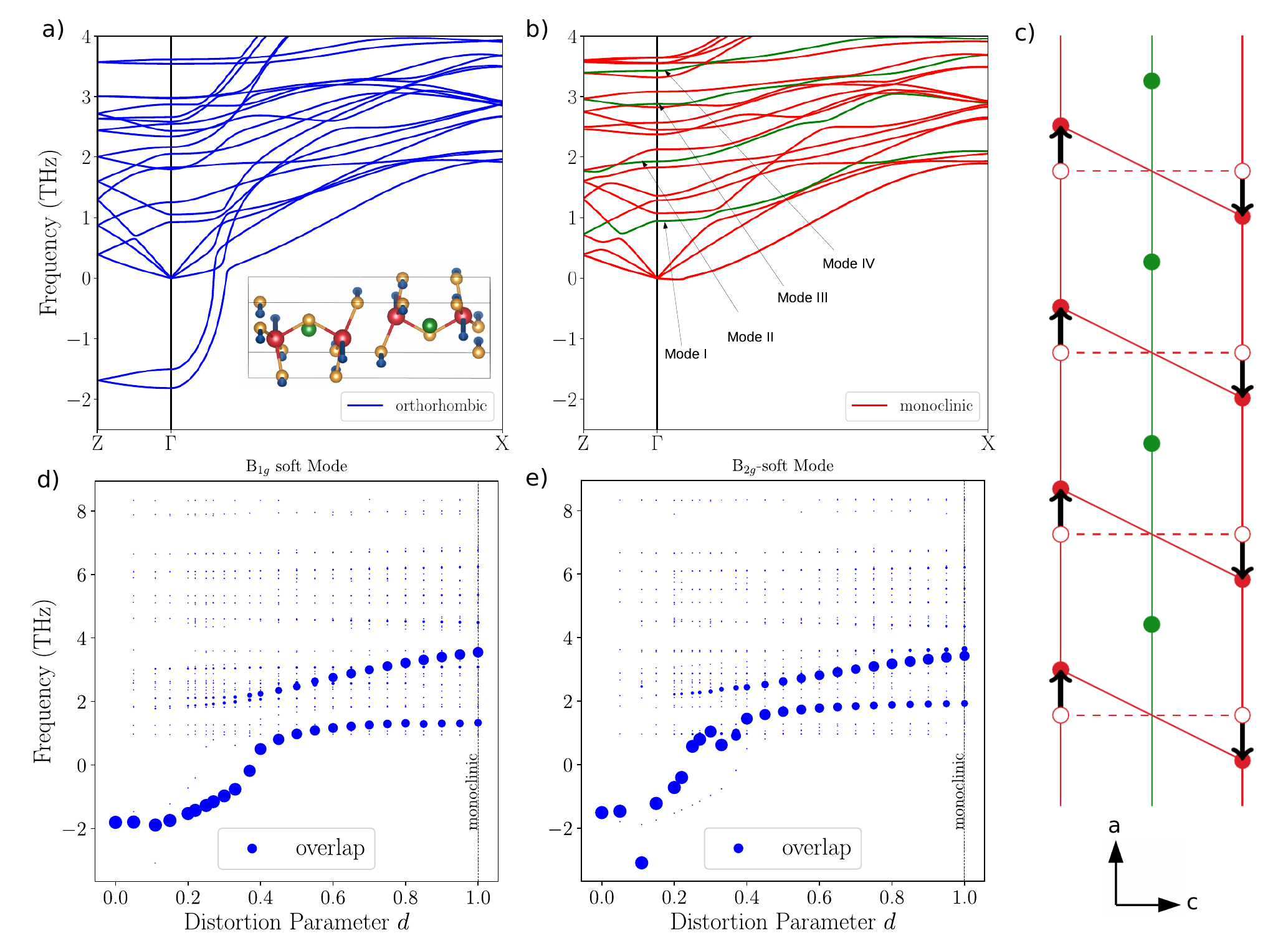}
     \caption{\textbf{Phonon properties of Ta$_2$NiSe$_5$:} a) TNSe Phonon dispersion using the vdW-optB88 functional for the orthorhombic phase. It shows two optical instabilities, one of which is the B$_{2 \text{g}}$ soft mode. The inset shows the eigenvector of this mode. b) Phonon dispersion of the monoclinic phase. The two unstable modes become stable. We highlighted the phonon branches of the four optical phonons I to IV for which the coupling to the electron bandstructure is discussed in section~\ref{sec:ep-coupling}.  The lowest acoustical phonon in the monoclinic phase shows an unnatural behaviour close $\Gamma$ which is an artefact of the numerical procedure applied.
     c) Schematic illustration of the shearing of the Tantalum atoms around the Nickel atoms, which has also been discussed in \cite{Kaneko2013, subedi2020} and is a key signature of the phase transition.
     d) and e) show the overlap of the unstable orthorhombic phonon eigenvectors with the phonon eigenvectors at different distortions $d$. The position of the blue dots show the phonon eigenenergy of the corresponding mode and the size the overlap between the mode eigenvector and the unstable orthorhombic mode B$_{1 \text{g}}$ mode d) and B$_{2 \text{g}}$ mode e) eigenvector. At $d=$0 we have the orthorhombic geometry and at $d=$1 the monoclinic geometry. One sees that the orthorhombic eigenvectors have maximal overlaps with the modes at 3.4 THz and 3.6 THz in the monoclinic phase.}
     \label{fig:tri-phonon}
 \end{figure*}

So far we have investigated the electronic degrees of freedom focusing on the effect of the lattice geometries. While we have introduced the concept of a structural instability in the prior sections we show how it is related to the lattice dynamics of the crystal.

The phonon dispersion of the orthorhombic as well as the monoclinic cell of TNSe are shown in Fig.~7a: the orthorhombic cell shows two unstable optical phonon modes around the $\Gamma$-point, characterized by imaginary frequencies (indicated as negative values here), while the monoclinic cell shows no phonon instabilities. 

The two unstable phonon modes belong to the B$_{1 \text{g}}$ and B$_{2 \text{g}}$ symmetry groups. 
%with the B$_{1 \text{g}}$ branch being the absolute minimum at $\Gamma$.
Both unstable phonons lead to monoclinic distortions and become stable in the monoclinic phase. The B$_{1 \text{g}}$ leads to a P2$_1$/m geometry and the B$_{2 \text{g}}$ mode leads to the C2/c symmetry. Just displacing the atoms within the unit cell and computing the total energy of the structure shows that the P2$_1$/m would be favored \cite{subedi2020}, however, also allowing for lattice distortions the C2/c symmetry is favored with an energy difference of 14.4 meV per unit cell. This is in agreement with the full relaxation as well as experimental findings observing a monoclinic C2/c symmetric geometry in the low-temperature phase of TNSe \cite{Sunshine1985, Nakano2018_2}. Therefore, we can identify the B$_{2 \text{g}}$ mode as the relevant structural instability which drives the phase transition. The eigenvector of this mode is displayed in the inset of Fig.~7a. It shows the characteristic monoclinic displacement along the chain direction which results in a shearing of the Ta atoms around the Nickel atom (see Fig.~7c) 
{which has been identified as the order parameter for the orthorhombic to monoclinic phase transition by Kaneko et al. \cite{Kaneko2013}}. 
Note, that a similar phonon analysis using ab-initio methods has been performed in Ref.\cite{subedi2020} and our calculations agree with this study.

For the sake of comparison we have performed a calculation of the phononic spectrum at the $\Gamma$-point for both geometries of TNS. We find one phononic instability in the orthorhombic structure which becomes stable in the monoclinic phase (see Supplementary Table 7). The instability has B$_{2 \text{g}}$ symmetry and drives, similarly to the TNSe case, the orthorhombic to monoclinic phase transition, which is consistent with the monoclinic groundstate being the energetically favoured one.

To investigate the evolution of the phonon instabilities when transforming the crystal from orthorhombic to monoclinic, we have performed calculations on increasingly distorted structures using the $d$ distortion parameter introduced in the previous section. Performing phonon calculations we obtained the phonon spectra and eigenvectors at the $\Gamma$-point for each of the distorted geometries. To follow the evolution of the phonon energies during the structural transition, we have computed the overlap between the orthorhombic eigenvectors of the two phonon instabilities with the phononic eigenvectors of all modes at a given value of the distortion parameter $d$. The result is displayed in Fig.~7c) and d) where the marker size is proportional to such overlaps. Note, that the phonon eigenvectors for a given distortion $d$ form an orthonormal system. Therefore, no overlap is visible at $d=$0 for the two unstable modes. As the structure is transformed to the monoclinic cell the different phonons start hybridizing which leads to non-zero overlap of the unstable orthorhombic phonons and the phonons at d$\neq$0. We observe that the two unstable phonons are being stabilized along the phase transition and hybridize predominantly into a pair of monoclinic phonons.  The unstable B$_{2\text{g}}$ has the strongest overlap with the monoclinic 3.4 THz and 1.9 THz phonon modes and the unstable B$_{1\text{g}}$ has the strongest overlap with the monoclinic 3.6 THz and 1.4 THz modes. In section \ref{sec:ep-coupling} we will show, that the two monoclinic B$_{2g}$ phonons couple strongly to the electronic degrees of freedom.

Further analysis of the phonon symmetries allows us to identify all three B$_{2 \text{g}}$ and eight A$_{\text{g}}$ phonon modes of the orthorhombic and all A$_{\text{g}}$ phonon modes of the monoclinic geometry. Their comparison with the phononic peak positions from recent experimental Raman measurements which are sensitive to B$_{2 \text{g}}$ and A$_{\text{g}}$ symmetry channel \cite{Kim2020, Blumberg2021, Blumberg2020, Blumberg2021_3} shows a good agreement of all monoclinic A$_{\text{g}}$ modes, except for one, and all orthorhombic A${_g}$ modes, even if our results are at T=0. Only the first two B$_{2\text{g}}$ modes, which are labelled 2 and 5 in the orthorhombic phase, and mode 5 in the monoclinic phase show a substantial energy difference between the calculated and the experimental results (see Fig.~8 and 9). In the most recent Raman experiments no soft mode is reported \cite{Kim2020,Blumberg2021, Kaiser2020_erratum, Blumberg2021_3, Blumberg2020} and the second B$_{2\text{g}}$ mode (labelled 5) differs by roughly 0.7 THz from the theoretically calculated value at T=0K. Note that K.Kim et al \cite{Kim2020} 
identify mode 5 as being almost degenerate with mode 3 in the orthorhombic phase and it evolves
into mode 5 of the monoclinic phase (see curve at 225K), where we also see a small discrepancy with our calculation. The two B$_{2\text{g}}$ modes exhibit a significant Fano-like broadening, while all other Raman active modes have a sharp lineshape. 
It is conceivable that the electronic structure methods used here do not properly capture all coupling channels of the phonon modes to the electronic degrees of freedom which, could be responsible for these effects. Another possible explanation are
inelastic phonon-phonon contributions due to anharmonic effects that are present for the two B$_{\text{2g}}$ modes. At the experimental temperature, phonon anharmonicity effects are expected to play a major role in renormalizing the phonon energies and stabilizing the unstable modes, as has been demonstrated in other materials exhibiting a soft-mode mediated phase transition~\cite{zhou2018}. This is especially relevant for the soft modes, which would explain why a very good agreement is observed for the other Raman modes even without including temperature effects. The anharmonic coupling of the soft mode to the continuum of other phononic modes could also be the source of the broadening of the Raman peak in the region just below 2~THz, which has been instead attributed to the coupling of the phonon to an excitonic instability. The idea of the fundamental role of phonon anharmonicity is supported by recent experimental Raman measurements on TNS by Ye et al Ref.~\cite{Blumberg2021}. Indeed it is shown that even though TNS undergoes a structural phase transition, no phonon softening is observed in the Raman data around the transition temperature at T=120K. As in the case of TNSe the authors claim that the phase transition is the result of an acoustic mode softening which in this case is coupled to a ferroelastic instability rather than an excitonic instability. 
In contrast our calculations do not show softening of the acoustic modes for both TNSe and TNS, but rather soft Raman active B$_{2g}$ modes (see Supplementary Figure 7).
Therefore, the lack of phonon softening in the Raman measurements in TNS is likely the result of the temperature stabilization of that  phonon, as explained above, and provides a further support to  the structural similarities between TNS and TNSe. A detailed investigation of the temperature effects in the phonon spectrum to fully corroborate this argumentation will be topic of future work. 
In the following section we investigate the coupling of such B$_{2\text{g}}$ modes to the electronic bandstructure using a frozen phonon approach. We found that especially phononic modes which displace the Tantalum and Nickel atoms along the chain direction of the crystal, such as the orthorhombic B$_{2\text{g}}$ modes, strongly modify the electronic bandstructure and hence could in turn be renormalized. 

 \begin{figure*}
     \centering
     \includegraphics[scale=0.42]{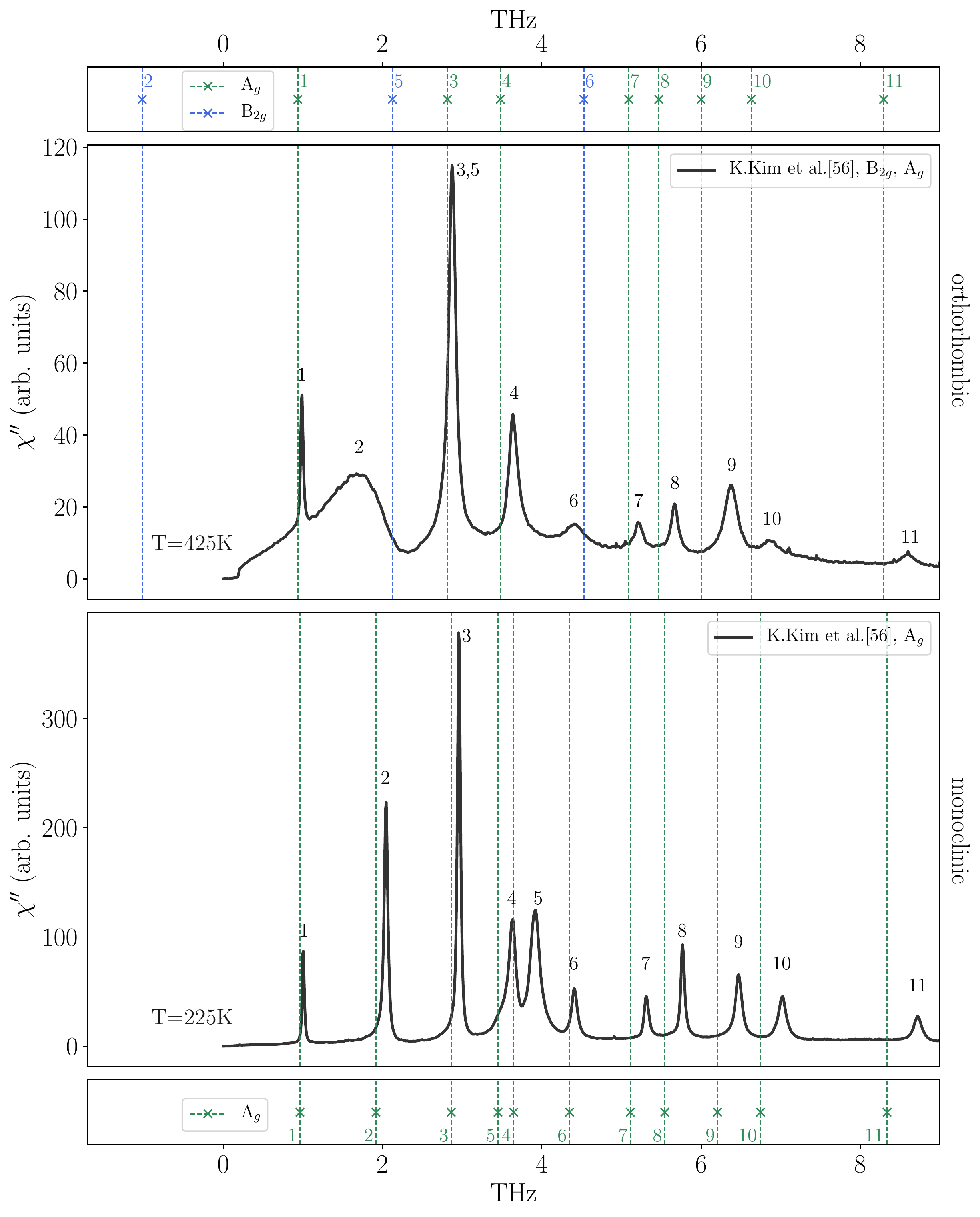}
     \caption{\textbf{Comparison of the Raman data from K.Kim et al. \cite{Kim2020} with our theory results:} Comparison of the theoretically calculated phonon eigenenergies at T=0K with the Raman spectra provided by K.Kim et al \cite{Kim2020}. The top panel shows the orthorhombic phonon spectrum and the bottom panel the monoclinic phonon spectrum. Note that peak 3 and 5 are almost degenerate in energy and peak 5 only appears in the $B_{2g}$-channel with a much smaller amplitude than peak 3 which appears in the A$_g$ channel. The theory spectra are obtained at T=0 K and are in good agreement with Raman spectra. Only the first two B$_{2\text{g}}$ modes of the orthorhombic geometry (peaks 2 and 5) differ substantially from their experimental values. This is consistent with these peaks having a significant broadening. The numerals in the experimental plots label the experimentally measured peaks by K.Kim\cite{Kim2020}. We assigned those numbers to the corresponding energies from our theory calculations by minimizing the energy differences and following the modes across the transition.}
     \label{fig: Raman comparison_Kim}
 \end{figure*}

 \begin{figure*}
     \centering
     \includegraphics[scale=0.42]{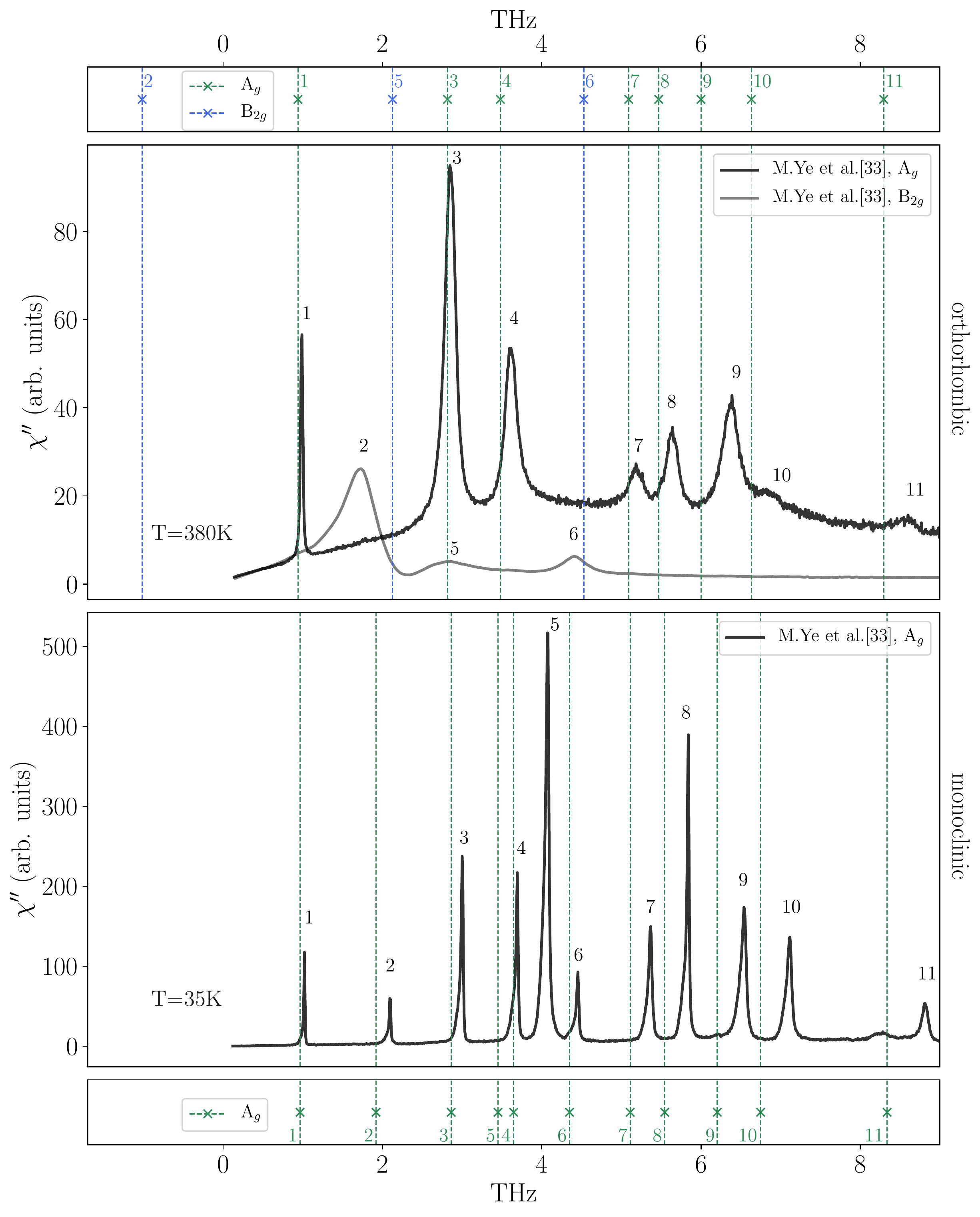}
     \caption{\textbf{Comparison of the Raman data from M.Ye et al. \cite{Blumberg2021} with the theory results:} Comparison of the theoretically calculated phonon eigenenergies at T=0K with the Raman spectra provided by M.Ye et al \cite{Blumberg2021}. The top panel shows the orthorhombic phonon spectrum and the bottom panel the monoclinic phonon spectrum on a linear scale. The theory spectra are obtained at T=0 K and are in good agreement with the Raman spectrum. Only the first two B$_{2\text{g}}$ modes of the orthorhombic geometry differ substantially from their experimental values. This is consistent with these peaks having a significant broadening. The numerals in the experimental plots label the experimentally measured peaks. We assigned those numbers to the corresponding energies from our theory calculations by minimizing the energy differences and following the modes across the transition.}
     \label{fig: Raman comparison_Blumberg}
 \end{figure*}

%clearpage
\subsection{\label{sec:ep-coupling} Electron-Phonon coupling}
We conclude our work by demonstrating the intimate coupling between the electronic and nuclear degrees of freedom by highlighting the influence of specific phononic modes on the electronic bandstructure. Time resolved ARPES measurements by Baldini et al. find that the gap presents oscillatory behaviour under photoexcitation \cite{Baldini2020}. The authors show that the momentum integrated photoelectron intensity is dominated by 4 frequency components: 0.98 THz, 2.11 THz, 3.0 THz and 3.67 THz. The four frequencies can be identified as the frequencies of Raman active phonons and using our phonon dispersion calculation (Fig.~7a) we can identify the corresponding modes, with Mode I at 1.1 THz, Mode II at 1.9 THz, mode III at 2.9 THz and IV at 3.4 THz. We observe that the modes II and IV have B$_{2 \text{g}}$ symmetry and show the characteristic shearing with respect to the Tantalum atoms around the Nickel atoms while the Modes I and III have A$_{\text{g}}$ symmetry do not exhibit this shearing. They can be identified with the peaks 1, 2, 3 and 5 in the Raman measurements of figures 8 and 9. Mode IV mode can also be identified as the monoclinic counterpart of the B$_{\text{2g}}$ soft-mode which is unstable in the orthorhombic geometry (see Fig.~7e).

To quantify how strongly the above discussed phonons couple to the electrons, we have calculated their effect on the electronic bandstructure by displacing the atoms along the phonon eigenmodes by an amount equal to the square root of their mean-squared displacement at zero temperature for both the positive and negative direction. The resulting bandstructures using the G$_0$W$_0$ method are shown in Fig.~10 (see Supplementary Figure 12 for the PBE results). We see that the Modes I and III show little influence whereas both B$_{2 \text{g}}$ modes have a strong impact on the electronic structure. We stress that both of these modes exhibit the characteristic shearing of the Ta atoms around the Nickel atom in chain direction, which we also observed during the structural phase transition (see Fig.~7c) . For the Mode IV the average gap between positive and negative displacement is larger than the equilibrium gap. Therefore, we expect a small increase of around $21$ meV of the bandgaps predicted in the previous section due to the coupling of this phonon to the electronic degrees of freedom. 

These results, beside confirming the strong influence of phonons on the electronic structure, hint towards the possibility of modulating the electronic bandgap through an external pumping of the above phononic modes. 

 \begin{figure*}[t]
     \centering
     \includegraphics[scale=0.55]{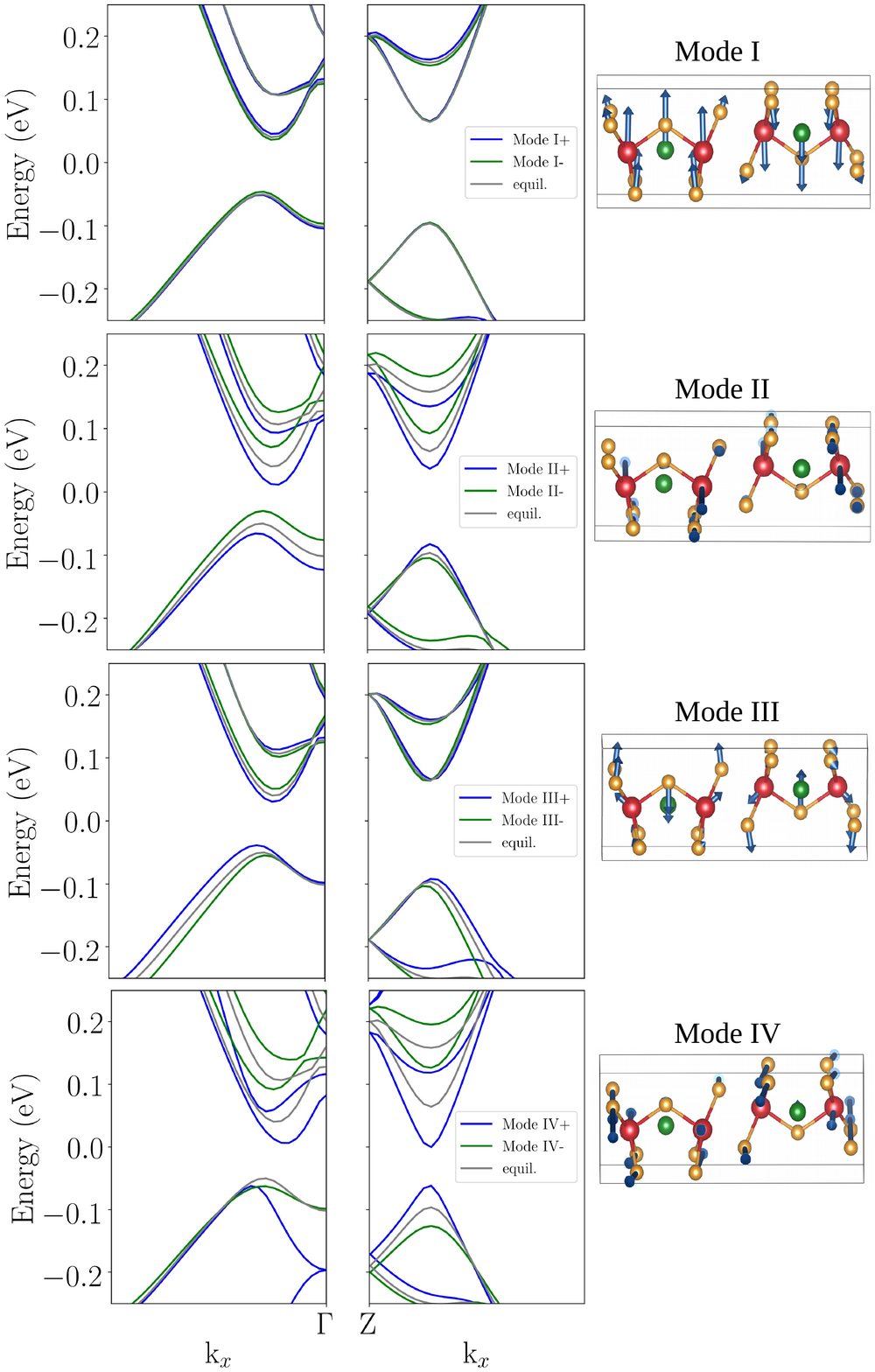}
     \caption{\textbf{Frozen phonon bandstructure:} PBE+G$_0$W$_0$ calculation using the mean-squared displacement at $T=0$~K along different phonon eigenvectors in both positive and negative direction. The bandstructure couples strongly to the B$_{\text{2g}}$ phonon modes II and IV at 2 THz and 3.4 THz which exhibit the characteristic shearing of the Ta atoms. The figure has been adapted from our work in ref.\cite{Baldini2020}.}
     \label{fig: PhononDisplacedBands}
 \end{figure*}

%clearpage
\section{\label{sec:conlusion} Discussion}
We have shown that the orthorhombic phase of both TNSe and TNS is unstable at low temperatures and exhibits a soft phonon mode, which signals the structural transition to the monoclinic phase.

Performing electronic bandstructure calculations using different exchange correlation functionals, as well as calculations using the G$_0$W$_0$ approximation, we have proven that in TNSe the structural transition drives the metal to semiconducting phase transition, with an indirect bandgap that is comparable to the experimental results \cite{Lu2017, Lee2019}, while in TNS the semiconductive electronic nature is preserved. 
This demonstrates that without the structural transition the experimentally observed gap opening at the critical temperature cannot be explained. At the same time our ab-initio results show that no excitonic effects are necessary to explain the experimental evidences.
Furthermore, during the structural transition a significant flattening of the valence bands is observed which is enhanced by electronic heating leading to almost flat bands which are in agreement with recently measured (pump-probe) ARPES signatures \cite{Baldini2020}. 

The electronic heating shows that the bandgap never closes for increasing electronic temperatures but shows the expected renormalization effects of standard semiconductors due to the increased carrier density. 

We discussed the possible origins of the structural transition in relation to the scenario of a 
purely electronic instability occurring at fixed ion positions and concluded that a structural 
distortion is required in order to observe a metal-insulator transition in TNSe. This is supported by scans of the total energy landscape and charge density as a function 
of the relevant distortion parameters which show no sign of an electronic instability.

Investigating the change of the electronic bandstructure for experimentally relevant phonon modes we found that the electrons tend to couple stronger to the phonon modes which displace along the chain direction and induce the characteristic shearing of the Ta atoms around the Ni atom. Especially the $3.4$~THz phonon mode is expected to lead to a small additional bandgap opening through the thermal occupation of this mode.
We conclude that, in the low temperature phase, TNSe and TNS behave as standard semiconductors.

\subsection*{Methods}
\subsubsection*{Computational Details}
All DFT and G$_0$W$_0$ calculations have been performed with the Vienna Ab initio Simulation Package (VASP) \cite{Vasp1,Vasp2,Vasp3,Vasp4,Vasp5} and in the case of phonon calculations we additionally used phonopy to compute phonon dispersions \cite{phonopy}. For the interpolation of the G$_0$W$_0$ bandstructure we have used the wannier90 package \cite{wannier90}.
All calculations have made use of the PBE pseudopotential set generated with VASP \cite{VaspPseudos}. \newline
The structural relaxations have been performed on a 24x4x6 k-mesh with an energy cutoff of 460 meV. A total free energy change of 1e-6 eV has been used as cutoff criterion which led to forces of 2.8 meV/angstrom per ion or less. The resulting input structures are shown in section 1 of the SI. 
The phonon dispersion calculation has been performed using the vdw-optB88 functional and a 4x2x1 supercell with a 4x4x3 k-mesh. The post-processing has been performed with the phonopy package \cite{phonopy}. The calculation of the phonon spectra at $\Gamma$ which are compared to the Raman data has been performed on a k48x4x6 mesh using a single unit cell. The calculation of the phonon overlaps as a function of the distortion parameter $d$ has been performed on a k16x8x3 mesh.
\newline
The electronic bandstructure calculations have been performed on a 24x8x6 k-mesh for the PBE functional and a 16x4x4 k-mesh for the modified Becke-Johnson and HSE hybrid functionals and the Hartree-Fock calculation. The mBJ-functional calculated self-consistently a c value of 1.2578 for the orthorhombic and a c-value of 1.2636 for the monoclinic phase of TNSe. For TNS c is 1.2620 for both structural phases. The bandstructures have been computed along the path M=$(\frac{\pi}{a}, 0 , \frac{\pi}{c})$, Z=$(0, 0, \frac{\pi}{c})$, $\Gamma$=$(0, 0, 0)$ and X=$(\frac{\pi}{a}, 0 , 0)$.%Convergence has been checked for all functionals.
\newline
The G$_0$W$_0$ calculations have been performed on a 12x4x2 k-mesh with 160 frequencies $\omega$ and 1160 states with a 100 meV energy Cutoff. The Wannierization has been performed with the Wannier90 package \cite{Mostofi2014}. A convergence study of the relevant parameters for the presented G$_0$W$_0$ calculation can be found in the Supplementary Note 7 and Supplementary Figures 13-15.

\subsection*{Data Availability}
The authors declare that the main data supporting the findings of this study are contained within the paper and its associated Supplementary Information. All other relevant data are available from the corresponding author upon reasonable request.

\subsection*{Code Availability}
The VASP software package is available at \url{https://www.vasp.at/}. The open-source software packages Wannier90 is available at \url{http://www.wannier.org/} under the GNU General Public License (v2). The open-source phonopy package is available at \url{https://phonopy.github.io/phonopy/} under the BSD license. 

\subsection*{Acknowledgements} 
We are grateful to E.Baldini, I.Mazin and Yann Gallais for enlightening discussions throughout the course of this work. We would like to thank M. Ye, G. Blumberg, K.Kim, B.J. Kim, M.J. Kim and S. Kaiser for sharing the experimental data of their Raman measurements and valuable discussions.\newline 
This work is supported by the European Research Council (ERC-2015-AdG-694097), Grupos Consolidados (IT1249-19) and the Flatiron Institute, a division of the Simons Foundation. We acknowledge funding by the Deutsche Forschungsgemeinschaft (DFG) under Germany’s Excellence Strategy - Cluster of Excellence Advanced Imaging of Matter (AIM) EXC 2056 - 390715994 and  by the Deutsche Forschungsgemeinschaft (DFG, German Research Foundation) – SFB-925 – project 170620586. Support by the Max Planck Institute - New York City Center for Non-Equilibrium Quantum Phenomena is acknowledged.. S. L. acknowledges support from the Alexander von Humboldt foundation.
G. M. acknowledges support of the Swiss National Science Foundation FNS/SNF through an Ambizione grant.

\subsection*{Author Contributions}
A.R and S.L designed the project. L.W performed the computations under the supervision of S.L. All authors discussed the computational results and contributed to the writing and revision of the manuscript.

\subsection*{Competing Interests}
The authors declare no competing interests.

\newpage
\clearpage 
\clearpage 
\clearpage 
\clearpage 
\clearpage 
\clearpage 
\clearpage 
\clearpage 

\section*{Supplementary Information}
\subsection*{Supplementary Note 1}
The bandstructures have been computed along the path M=$(\frac{\pi}{a}, 0 , \frac{\pi}{c})$, Z=$(0, 0, \frac{\pi}{c})$, $\Gamma$=$(0, 0, 0)$ and X=$(\frac{\pi}{a}, 0 , 0)$.

\subsection*{\label{A: TotEnergies}Supplementary Note 2: Total energy minimum with different functionals}
To track the total energy changes during the relaxation for different functionals we have computed the energy of the system along the structural transition. We employed again the $d$ parameter introduced in section 2.4 of the main text to parameterize the transition. The resulting total energy vs $d$ distortion graph is shown in 
Supplementary Figure~\ref{fig: totEnergy}. For all functionals the total energy is displayed with the orthorhombic value set 0 eV. The figure shows, that using the PBE functional has the energy minimum at slightly lower values for the distortion and the HSE has its energy minimum at stronger distortion values of $d$. Nevertheless, a monoclinic structure is energetically more stable than the orthorhombic geometry for all investigated functionals. 

\subsection*{Supplementary Note 3: Spin-Orbit Coupling}
 The results for the PBE, mBJ and HSE03 with and without Spin-Orbit Coupling (SOC) functional are shown in the Supplementary Figures \ref{fig: SOC_PBE}, \ref{fig: SOC_mBJ} and \ref{fig: SOC_HSE}. They show that SOC leads to a bandsplitting which is most apparent on the M-Z path for all band dispersions. Note, that this band splitting is not a splitting of spin up and down states. Near the bandedge the spin orbit coupling has just a small influence on the bandstructures. Only for the monoclinic geometry of TNSe using the PBE functional changes are apparent with the bandgap halving and the becoming and direct at $\Gamma$. A detailed comparison of the obtained bandgaps is presented in Supplementary Table \ref{tab: gapSOC}. We conclude, that the effects of SOC are negligible for our discussion in the main text. These results are in agreement with a prior study by Sugimoto et al. \cite{Sugimoto2018}.
 
 \subsection*{\label{A: exchange}Supplementary Note 4: Increasing the exchange-correlation contribution in the functionals}
Increasing the exchange and correlation contribution considered in either the HSE hybrid functionals or the modified-Becke-Johnson functional will lead to a bandgap opening even in the orthorhombic unit cell. In the modified Becke-Johnson (mBJ) functional the amount of exchange and correlation interaction included is controlled by the c parameter. It controls the amount of Becke-Roussel exchange \cite{Becke2006,Becke1989}, which approximates exact exchange effects via an effective potential, plus a screening term \cite{Tran2009}. In the HSE hybrid functionals the exact exchange contribution is calculated directly using the Kohn-Sham orbitals during the self consistent iterations. Parts of the PBE exchange interaction are replaced by the exact exchange interaction \cite{Heyd2003}. The admixture of exact and PBE exchange in the functional is controlled by the mixing parameter $\alpha$ which is commonly set to 0.25 for the HSE functionals. Varying it we can modify the amount of exchange interaction analogously to varying the $c$ parameter in the mBJ functional.  
The results are shown in the Supplementary Figure~\ref{fig: cEvolution} and Supplementary Figure~\ref{fig: alphaEvolution}. We see, that for values of $c=$1.6 or higher a bandgap opens, while the system is still metallic for the self consistently calculated c-Value of 1.26. Analogously we can control the amount of exchange and correlation included in the range separated hybrid functionals. To test this we have varied the mixing parameter $\alpha$, while keeping the range separation parameter at 0.3. This way $\alpha=0.25$ reproduces the result for the HSE03 functional. The resulting bandstructures can be seen in Supplementary Figure~\ref{fig: alphaEvolution}. The behaviour is similar to the mBJ case, with the orthorhombic cell becoming a semiconductor for values of $\alpha > 0.45$.

\subsection*{\label{A: Inf.Sym.Breaking}Supplementary Note 5: Minimally symmetry broken electron dispersion}
 We used the displacement parameter $d$ as defined in section 3 of the main text to introduce a minimal distortion to the lattice (d=0.05). This procedure breaks the relevant orthorhombic lattice symmetries. If the phase transition of TNSe is fully electronic the breaking of the symmetries should already induce the metal to semiconductor transition. The result for the electronic dispersion of the symmetry broken geometry is shown in Supplementary Figure~\ref{fig: smallDistortion}. We see that all bandstructures ,independent of the exchange correlation functional, do not exhibit a metal to insulator transition or a significant gap opening. In fact they agree very well with the exact orthorhombic results.

\subsection*{\label{A: STS-Gap-Corr}Supplementary Note 6: STS Gap Correction}
In STS measurements a metal to semiconductor transition has been reported at the critical temperature for TNSe \cite{Lee2019}. The reported bandgap is 300 meV and is significantly bigger than the bandgap obtained from optics measurements of 160 meV \cite{Lu2017}. The discrepancy can be explained taking the different geometries of the investigated samples of TNSe in account. The sample used for the STS measurement exhibits an $\beta$-angle of 92.5$^\circ$ and is such significantly bigger than the literature result of 90.5$^\circ$-90.7$^\circ$ \cite{Sunshine1985,Nakano2018}. To correct the bandgap taking this discrepancy into account, we computed the bandstructures for both our monoclinic unit cell as well as a unit cell with 92.5$^\circ$ $\beta$-angle. The obtained bandgaps are 40meV for our relaxed geometry and 66 meV for the 92.5$^\circ$ geometry using the PBE functional. Taking the ratio of these two and multiplying the 300 meV experimental gap, we obtain a correction of the STS gap to 181 meV.

\subsection*{\label{A: GW Convergence}Supplementary Note 7: Convergence of the G$_0$W$_0$ calculations}

We present the details of the convergence of the G$_0$W$_0$ calculation for the monoclinic unit cell. To display the convergence behaviour of the calculation we performed convergence studies for the number of k-Points needed in each direction, the number of unoccupied bands and the energy cutoff we have to consider for the response function calculation and size of the frequency grid of the frequency integration. It is important to note, that the energy cutoff for the response function and the number of unoccupied states are not independent convergence parameters and thus, we use the standard method and converge them simultaneously \cite{Klimes2014}. The bandstructures have been obtained via post-processing using the Wannier90 package \cite{wannier90}. \newline 
For the presented convergence studies we have employed the Density Functional Theory solution using the PBE functional as starting point. A discussion of the effect using different functionals such as the HSE hybrid functional is in the section 2.3 of the main text.

For the bandstructure convergence of TNSe with varying k-mesh we will assume that the spatial directions for the k-mesh converge independently and perform a series of calculations with varying grid size in all three directions. To measure the convergence of the bandstructure we will compute the fundamental bandgap and extrapolate using a function $f(x)=a+b/(x+c)$ with $a$,$b$ and $c$ as fitting parameters. For these k-mesh calculations we used 640 unoccupied states with a 80eV Cutoff for the response function and 160 frequency grid points in the calculation of the screened interaction. The k-meshes are $k_x\times4\times2$, $12\times k_y\times4$ and $12\times4\times k_z$ with $k_x, k_y$ and $k_z$ are varied. The result is shown in Supplementary Figure~\ref{fig: kConvergence}. One sees, that the convergence in $k_y$ and $k_z$ along the corresponding reciprocal lattice directions converges quite fast. This is different for $k_x$ as the bandstructure is highly dispersive in x-direction and many grid-points are needed to sample it accurately.   

We also performed a simultaneous convergence of the energy cutoff for the response function and the number of unoccupied states. Convergence against the exact result should be in first order proportional to 1/Number of unoccupied states \cite{Klimes2014}. We have performed the convergence calculation in the standard way increasing response function cutoff and the number of included orbitals simultaneously. The given combinations are shown in Supplementary Table~\ref{tab: eCut}.
The result of the energy convergence is presented in Supplementary Figure~\ref{fig: eCut}. We see, that including 1160 unoccupied states leads to a quite well converged bandgap already.

Lastly we also investigate of the convergence with increasing number of frequency grid points. This time the test setup is a k-mesh of 12x4x2 with 880 unoccupied states and a 100eV Cutoff. The result is shown in Fig~\ref{fig: nOmega}. We see, that full convergence is only achieved for a large number of frequencies considered due to the small bandgap. 

\subsection*{\label{A: BSE-Convergence}Supplementary Note 8: BSE Convergence}
We also investigated the convergence behaviour of the solution of the Bethe-Salpether equation \cite{Salpeter1951,Sander2015} for the monoclinic phase of TNSe. We performed the BSE using the eigenenergies and screened interaction from the G$_0$W$_0$ calculation. The calculations show, that for all convergence parameters the excitonic binding energy decreases proportional to the bandgap. This together with the fact, that even more converged G$_0$W$_0$ calculations in k$_x$ tend to increase the G$_0$W$_0$ bandgap suggests, that for the material the excitonic binding energy does not exceed the bandgap. A detailed analysis of all relevant convergence parameters follows:   

The solution of the Bethe-Salpether equation is very sensitive to the k-mesh. Thus, we investigated the solution of the BSE with an increasing k-mesh along the three different axes. We use a test setup with 160 frequencies for the calculation of the screened interaction, 640 unoccupied states (80eV cutoff) during the G$_0$W$_0$ calculation and include again the first 12 valence and 14 conduction bands in the BSE. The k-meshes used are k$_x$x4x2, 8xk$_y$x2 and 8x4xk$_z$, where k$_x$, k$_y$ and k$_z$ are variable and are be successively increased. The results are depicted in Supplementary Figure~\ref{fig: BSE_kConvergence}. 
We see, that while the dielectric function converges quickly with an increasing k-mesh in y-direction and z-direction, we need many k-points in k$_x$ direction to obtain we reasonable result. The reason is, that the bandgap minimum is shifted slightly toward x-direction after the G$_0$W$_0$ calculation and therefore only very dense k-meshes can sample the highly dispersive energies in x-direction. 

Secondly, we investigate the effect onto the BSE calculation for an increasing cutoff in the G$_0$W$_0$ calculation and an increasing number of unoccupied states. The test setup is working with a 12x4x2 k-mesh, 160 frequencies for the calculation of the screened interaction and including the first 12 valence and 14 conduction band states in the BSE. We employed the Tamm-Dancoff approximation \cite{Dancoff1950,Sander2015}.
The results are depicted in Supplementary Figure~\ref{fig: BSE_CutoffConvergence}. Using 880 unoccupied states with a 80eV cutoff already gives a well converged dielectric function and excitonic binding energies.

The third convergence parameter we investigate is the effect of including an increasing number of frequencies in the computation of the screened interaction onto the result of the Bethe-Salpeter equation. The test setup is using a 12x4x2 k-mesh, 640 unoccupied states and the first 12 valence and 14 conduction band states in the BSE. The results is depicted in Supplementary Figure~\ref{fig: BSE_omegaConvergence}. We see, that the BSE solution converges quite fast with the number of frequencies included in the calculation of the screened interaction. The dielectric function only obtains a rigid shift, which can be explained by the decreasing bandgap of the underlying G$_0$W$_0$ bandstructure calculation. 

Lastly, we show the BSE convergence properties with an increasing number of valence and conduction bands included in the BSE calculation. The test setup includes a 12x4x2 k-mesh, 880 unoccupied states, a 80 eV cutoff and 160 frequencies in the calculation of the screened interaction. The number of valence (v) and conduction bands (c) included is varied. The convergence of the dielectric function and the eigenvalues of the BSE solution is displayed in the Supplementary Figure~\ref{fig: BSE_nVC}. We see, that we always obtain well converged eigenvalues, but need to include at least the first 12 valence and the first 14 conduction bands around the Fermi level to obtain a fully converged dielectric function. The excitonic eigenvalues are already very well converged only the first valence and conduction bands.

\begin{table}[ht]
    \centering
    \begin{tabular}{l|c|c|c}
     & a (in $\SI{}{\angstrom}$) & b (in $\SI{}{\angstrom}$) & c (in $\SI{}{\angstrom}$) \\ \hline
        Experiment \cite{Nakano2018} & 3.492 & 12.814 & 15.649 \\
        vdW-optB88 & 3.517 & 12.982 & 15.776 \\
        vdW-optPBE & 3.532 & 13.325 & 15.845 \\
        PBE        & 3.510 & 14.160 & 15.776 \\
    \end{tabular}
    \caption{Lattice parameters after full relaxation of the TNSe compound in the monoclinic phase using different functionals. The experimental reference data has been measured at 30K via X-ray diffraction \cite{Nakano2018}. The theory calculations are performed at T=0K.}
    \label{tab:lattice params}
\end{table}

\begin{table}[ht]
    \centering
    \begin{tabular}{l|c|c|c}
     & a (in $\SI{}{\angstrom}$) & b (in $\SI{}{\angstrom}$) & c (in $\SI{}{\angstrom}$) \\ \hline
        Experiment\cite{Nakano2018_2} & 3.503 & 12.870 & 15.677 \\
        vdW-optB88 & 3.512 & 12.993 & 15.771 \\
        vdW-optPBE & 3.526 & 13.252 & 15.834 \\
        PBE        & 3.504 & 14.190 & 15.762 \\
    \end{tabular}
    \caption{Lattice parameters after relaxation of the TNSe compound with enforced orthorhombic symmetry. The experimental values have been obtained via X-ray diffraction at T=400K \cite{Nakano2018_2}. The theory calculations are performed at T=0K.}
    \label{tab:lattice params ortho}
\end{table}

\begin{table}[ht]
    \centering
    \begin{tabular}{l|c|c|c}
     & a (in $\SI{}{\angstrom}$) & b (in $\SI{}{\angstrom}$) & c (in $\SI{}{\angstrom}$) \\ \hline
        Experiment\cite{Sunshine1985}    & 3.415 & 12.146 & 15.097 \\
        orthorhombic  & 3.430 & 12.200 & 15.203 \\
        monoclinic    & 3.428 & 12.223 & 15.203 \\
    \end{tabular}
    \caption{Lattice parameters for TNS of both the orthorhombic phase and the monoclinic phase obtained via full relaxation. The experimental values have been measured via X-ray diffraction at T=278 K. \cite{Sunshine1985}.}
    \label{tab:lattice params TNS}
\end{table}

\newpage
 
\begin{table}[ht]
    \centering
    \begin{tabular}{l|l|c|c|c}
lattice vector & x (in $\SI{}{\angstrom}$) & y (in $\SI{}{\angstrom}$) & z (in $\SI{}{\angstrom}$) \\ \hline
  &            &             &             \\
a & 3.51177786 & 0.00000000  &  0.00000000 \\
b & 1.75588893 & 6.49659343  &  0.00000000 \\ 
c & 0.00000000 & 0.00000000  & 15.77141641 \\ 
  &            &             &             \\    
  &            &             &             \\ 
 species & a  & b  & c                        \\ \hline
Ta &  0.22143275 & 0.55713450 & 0.88896088\\
Ta &  0.22143275 & 0.55713450 & 0.61103912\\
Ta &  0.77856725 & 0.44286550 & 0.11103912\\
Ta &  0.77856725 & 0.44286550 & 0.38896088\\
Ni &  0.70190352 & 0.59619297 & 0.75000000\\
Ni &  0.29809648 & 0.40380703 & 0.25000000\\
Se &  0.58116235 & 0.83767530 & 0.86174128\\
Se &  0.58116235 & 0.83767530 & 0.63825872\\
Se &  0.41883765 & 0.16232470 & 0.13825872\\
Se &  0.41883765 & 0.16232470 & 0.36174128\\
Se &  0.14695064 & 0.70609873 & 0.04873952\\
Se &  0.14695064 & 0.70609873 & 0.45126048\\
Se &  0.85304936 & 0.29390127 & 0.95126048\\
Se &  0.85304936 & 0.29390127 & 0.54873952\\
Se &  0.32729988 & 0.34540023 & 0.75000000\\
Se &  0.67270012 & 0.65459977 & 0.25000000\\
   \end{tabular}
    \caption{Result of the relaxation of the orthorhombic cell with Cmcm symmetry obtained using the vdw-optB88 functional. The atoms are given in units of the lattice vectors a,b and c}
    \label{tab:ortho cell}
\end{table}

\begin{table}[ht]
    \centering
    \begin{tabular}{l|l|c|c|c}
lattice vector & x (in $\SI{}{\angstrom}$) & y (in $\SI{}{\angstrom}$) & z (in $\SI{}{\angstrom}$) \\ \hline
&                       &                       &                 \\
a  &      3.5172791481  &       0.0000000000    &     0.0000000000\\
b  &      1.7642919084  &       6.4903780477    &     0.0000000000\\
c  &     -0.1695581062  &      -0.0018790259    &    15.7763685968\\
  &            &             &             \\    
  &            &             &             \\
 species & a  & b  & c                        \\ \hline
Ta  &   0.788009744     &    0.446403261    &     0.888662426\\
Ta  &   0.765747213     &    0.446396149    &     0.611343122\\
Ta  &   0.209582861     &    0.558275853    &     0.111357977\\
Ta  &   0.231832330     &    0.558274801    &     0.388645515\\
Ni  &   0.297194687     &    0.407655269    &     0.750000961\\
Ni  &   0.701051209     &    0.596787458    &     0.250006008\\
Se  &   0.411119367     &    0.165767233    &     0.861499543\\
Se  &   0.424071823     &    0.165757101    &     0.638461393\\
Se  &   0.586395710     &    0.838882702    &     0.138539629\\
Se  &   0.573462842     &    0.838897364    &     0.361493392\\
Se  &   0.859540334     &    0.295478962    &     0.048418035\\
Se  &   0.845322448     &    0.295464258    &     0.451556475\\
Se  &   0.137960710     &    0.709232641    &     0.951592761\\
Se  &   0.152210287     &    0.709218409    &     0.548427477\\
Se  &   0.670947177     &    0.658753928    &     0.750003802\\
Se  &   0.327025141     &    0.345806896    &     0.249991137\\
    \end{tabular}
    \caption{Result of the relaxation of the triclinic cell using the vdw-optB88 functional. The atomic positions are given in units of the lattice vectors a,b, and c.}
    \label{tab:mono cell}
\end{table}

\begin{table}
    \centering
    \begin{tabular}{c|c|c|c|c|c}
               & TNSe, ortho & TNSe, mono &  & TNSe, ortho,SOC & TNSe, mono, SOC  \\ \hline
               &             &            &  &                 &                  \\
         PBE   &  metallic   &  40 meV    &  &  metallic       & 18 meV (-55\%)          \\
         mBJ   &  metallic   &  101 meV   &  &  metallic       & 120 meV (+18\%)         \\
         HSE03 &  metallic   &  183 meV   &  &  metallic       & 179 meV (-2\%)          \\
               &             &            &  &                 &                  \\
               &             &            &  &                 &                  \\               
               & TNS, ortho & TNS, mono &  & TNS, ortho,SOC & TNS, mono, SOC  \\ \hline
               &             &            &  &                 &                  \\
        PBE    &  metallic   &   102 meV  &  &   metallic      & 102 meV (+0\%)         \\
        mBJ    &  42 meV     &   151 meV  &  &   98 meV (+133\%)       & 178 meV (+18\%)         \\
        HSE03  &  248 meV    &   352 meV  &  &   250 meV (+1\%)      & 362 meV (+3\%)         \\
    \end{tabular}
    \caption{Comparison of the obtained bandgaps for TNSe and TNS with various functionals including SOC and neglecting it. We obtain quite good agreement between the SOC result and the non SOC result except for TNSe in the monoclinic phase for the PBE functional.}
    \label{tab: gapSOC}
\end{table}

\begin{table}%[h!]
    \centering
    \scalebox{0.99}{
    \begin{tabular}{c|c|c|}
         Mode & Orthorhombic (THz) & Monoclinic (THz) \\ \hline
  1  &   -2.581   &   0.000  \\ 
  2  &    0.000   &   0.000  \\ 
  3  &    0.000   &   0.000  \\ 
  4  &    0.000   &   1.156  \\ 
  5  &    0.761   &   1.224  \\ 
  6  &    1.160   &   1.543  \\ 
  7  &    1.224   &   1.703  \\ 
  8  &    1.670   &   1.969  \\ 
  9  &    1.890   &   2.084  \\ 
 10  &    2.580   &   2.668  \\ 
 11  &    2.693   &   3.117  \\ 
 12  &    3.107   &   3.331  \\ 
 13  &    3.136   &   3.403  \\ 
 14  &    3.299   &   3.628  \\ 
 15  &    3.370   &   3.775  \\ 
 16  &    3.430   &   4.160  \\ 
 17  &    3.567   &   4.255  \\ 
 18  &    3.920   &   4.397  \\ 
 19  &    4.255   &   4.679  \\ 
 20  &    4.393   &   4.966  \\ 
 21  &    4.578   &   5.382  \\ 
 22  &    4.694   &   5.675  \\ 
 23  &    6.410   &   6.400  \\ 
 24  &    7.338   &   7.159  \\ 
 25  &    7.657   &   7.403  \\ 
 26  &    7.701   &   7.476  \\ 
 27  &    7.785   &   7.520  \\ 
 28  &    7.817   &   7.585  \\ 
 29  &    7.823   &   8.025  \\ 
 30  &    7.911   &   8.029  \\ 
 31  &    7.967   &   8.093  \\ 
 32  &    8.170   &   8.134  \\ 
 33  &    8.183   &   8.259  \\ 
 34  &    8.214   &   8.439  \\ 
 35  &    8.490   &   8.495  \\ 
 36  &    8.668   &   8.686  \\ 
 37  &    8.733   &   8.777  \\ 
 38  &    9.000   &   9.311  \\ 
 39  &    9.313   &   9.539  \\ 
 40  &    9.574   &   9.715  \\ 
 41  &    9.691   &   9.765  \\ 
 42  &    9.960   &   9.958  \\ 
 43  &   10.142   &  10.187  \\ 
 44  &   10.174   &  10.242  \\ 
 45  &   11.362   &  11.451  \\ 
 46  &   11.397   &  11.459  \\ 
 47  &   11.477   &  11.544  \\ 
 48  &   11.536   &  11.571  \\
    \end{tabular}
    }
    \caption{Eigenvalues of the phononic modes of TNS calculated at the $\Gamma$-Point.}
    \label{tab:phonons_TNS}
\end{table}

\begin{table}%[ht]
    \centering
    \begin{tabular}{l|l}
     cutoff (eV) & orbitals \\ \hline
        60  & 640   \\
        80  & 880   \\
        100 & 1160   \\
        120 & 1480   \\
    \end{tabular}
    \caption{Energy and orbital cutoff pairs used to check the convergence of the response function in the G$_0$W$_0$ calculation.}
    \label{tab: eCut}
\end{table}

\begin{figure}%[h!]
    \centering
    \includegraphics[scale=0.4]{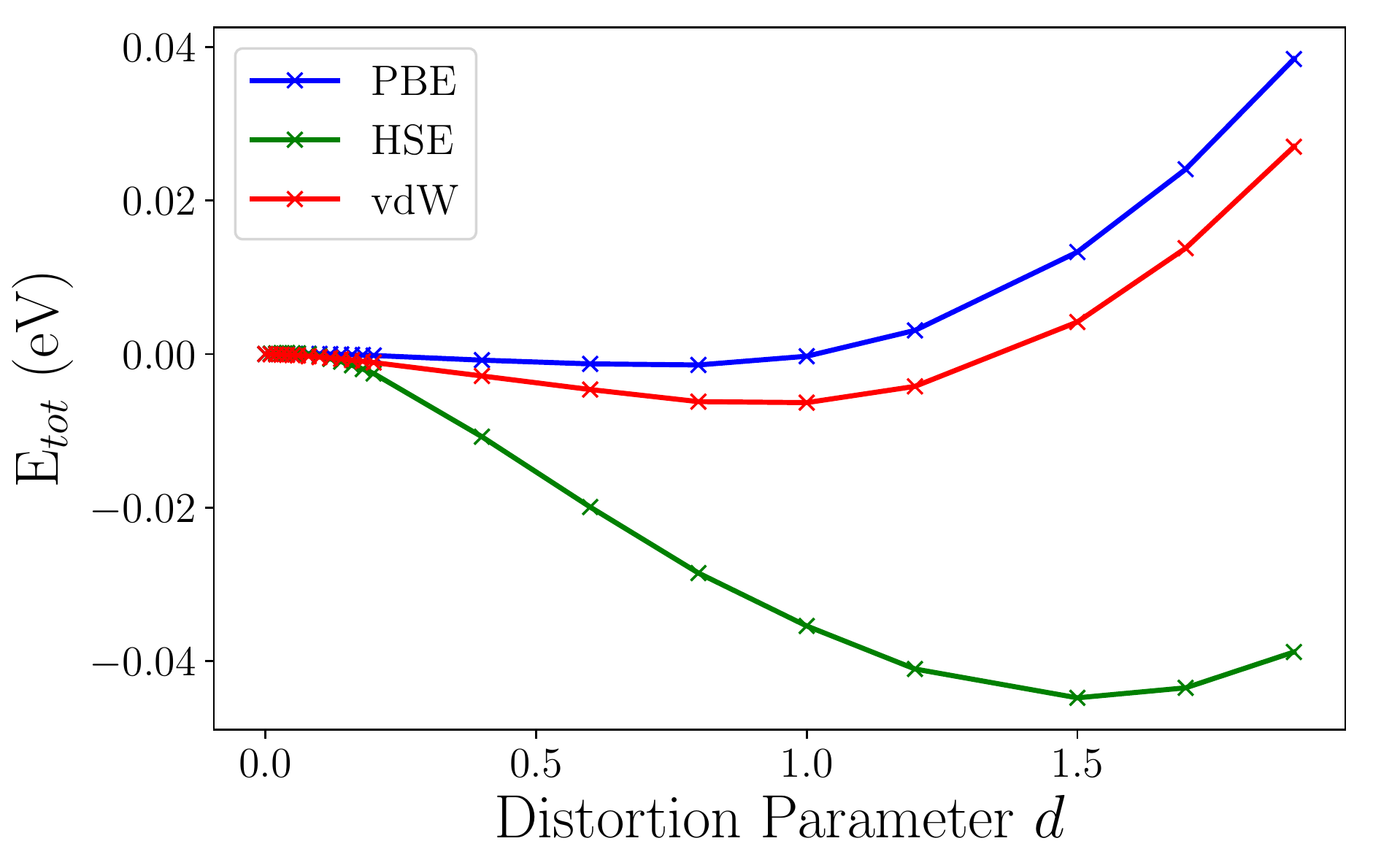}
    \caption{Total energy calculations along the structural transition parameterized by the distortion parameter $d$. For all functionals their total energy change has been computed as difference to its orthorhombic value}
    \label{fig: totEnergy}
\end{figure}

\clearpage

\begin{figure*}%[h!]
    \centering
    \includegraphics[scale=0.4]{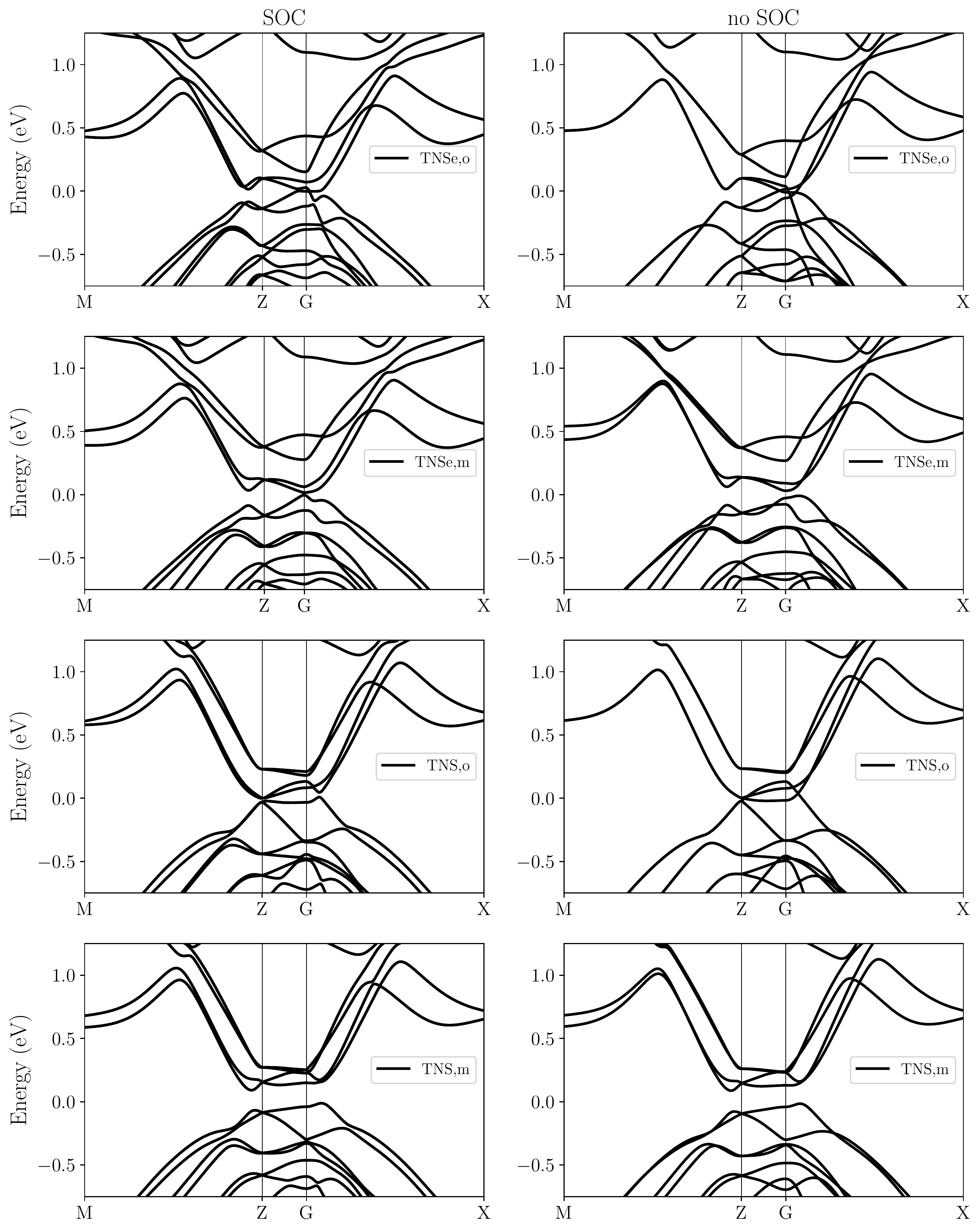}
    \caption{Here we compare the bandstructures obtained with the PBE functional including spin orbit coupling and neglecting it for both structural phases of TNSe and TNS. The letters o and m describe the orthorhombic and monoclinic geometry in the plot labels.}
    \label{fig: SOC_PBE}
\end{figure*}

\begin{figure*}%[h!]
    \centering
    \includegraphics[scale=0.4]{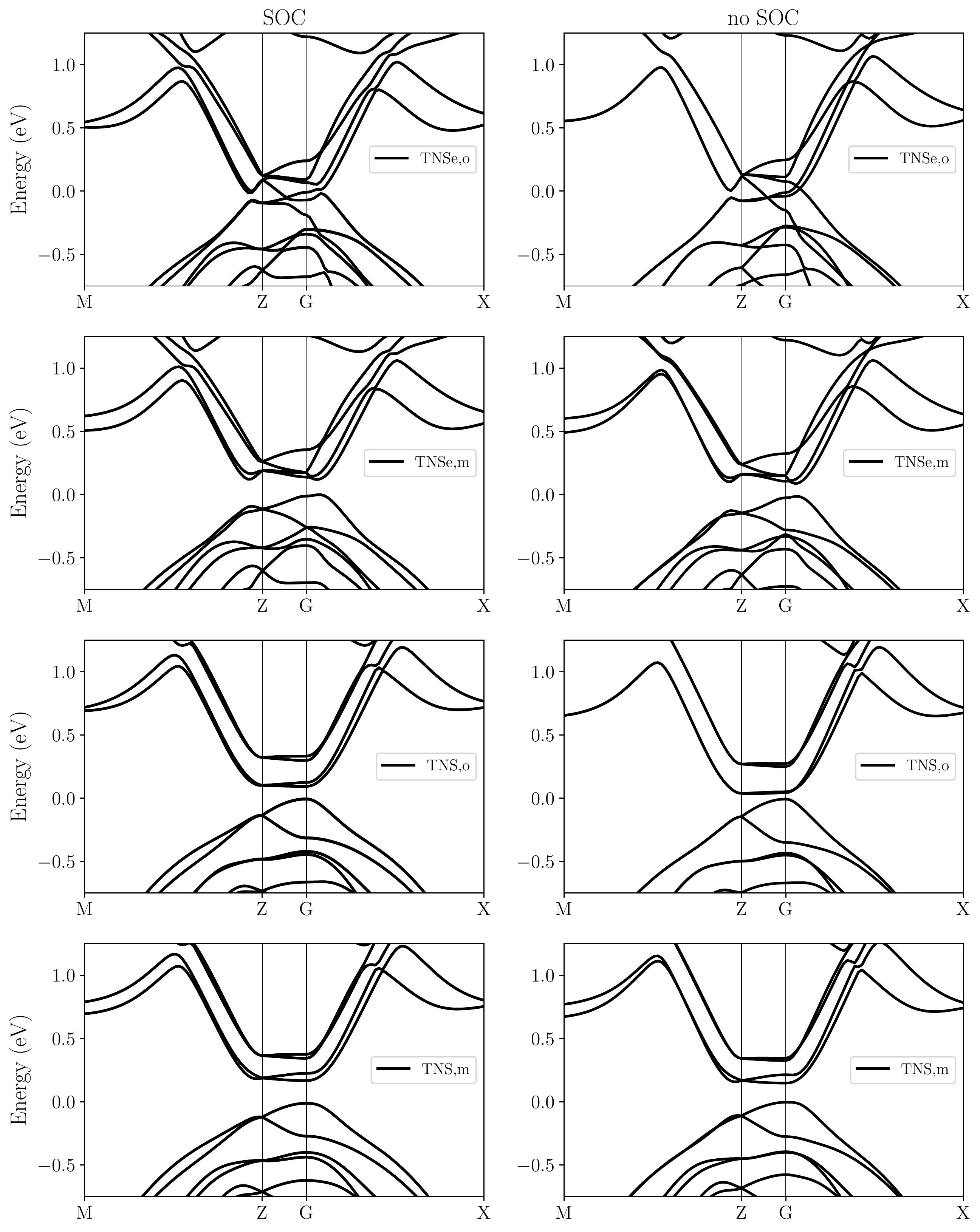}
    \caption{Here we compare the bandstructures obtained with the mBJ functional including spin orbit coupling and neglecting it for both structural phases of TNSe and TNS. The letters o and m describe the orthorhombic and monoclinic geometry in the plot labels.}
    \label{fig: SOC_mBJ}
\end{figure*}

\begin{figure*}%[h!]
    \centering
    \includegraphics[scale=0.4]{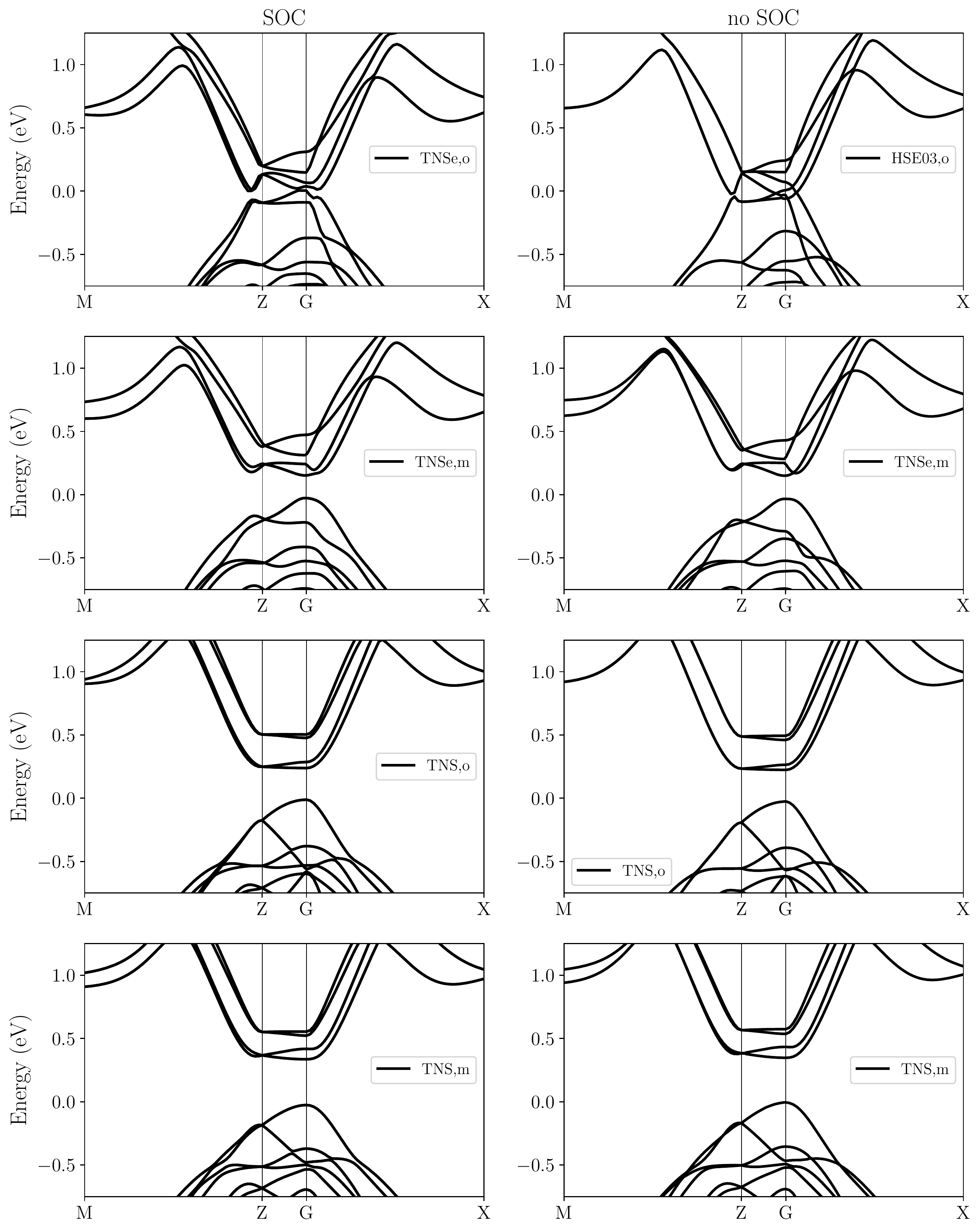}
    \caption{Here we compare the bandstructures obtained with the HSE03 functional including spin orbit coupling and neglecting it for both structural phases of TNSe and TNS. The letters o and m describe the orthorhombic and monoclinic geometry in the plot labels.} 
    \label{fig: SOC_HSE}
\end{figure*}

\clearpage
\clearpage

\clearpage

\begin{figure*}%[h!]
    \centering
    \includegraphics[scale=0.4]{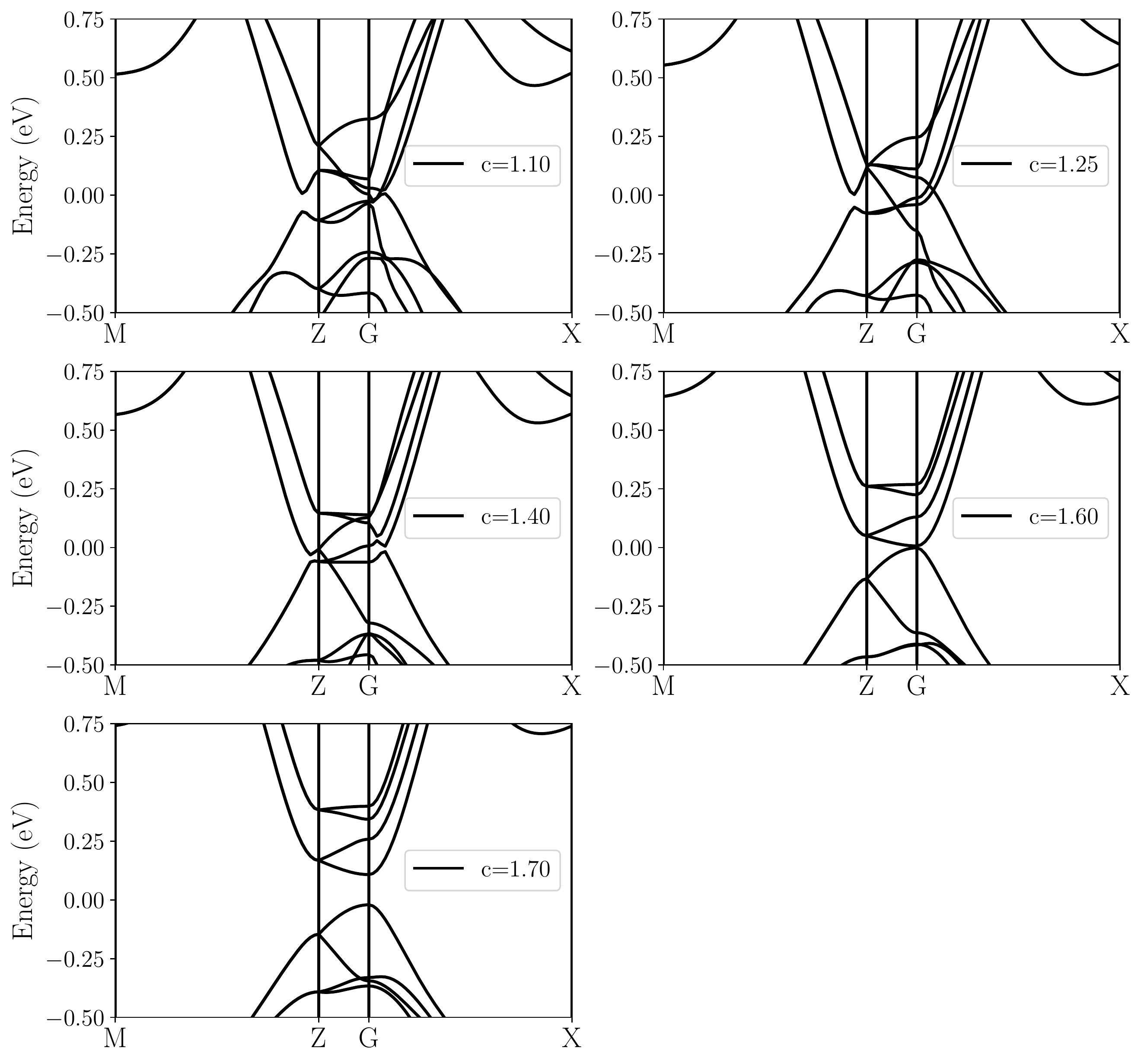}
    \caption{Bandgap opening as we increase the amount of exchange considered in the mBJ-functional by increasing the c-parameter in the orthorhombic phase. For values higher than 1.6 a gap opens. The self consistently calculated c-Value is 1.26.}
    \label{fig: cEvolution}
\end{figure*}

%\newpage

\begin{figure*}%[h!]
    \centering
    \includegraphics[scale=0.35]{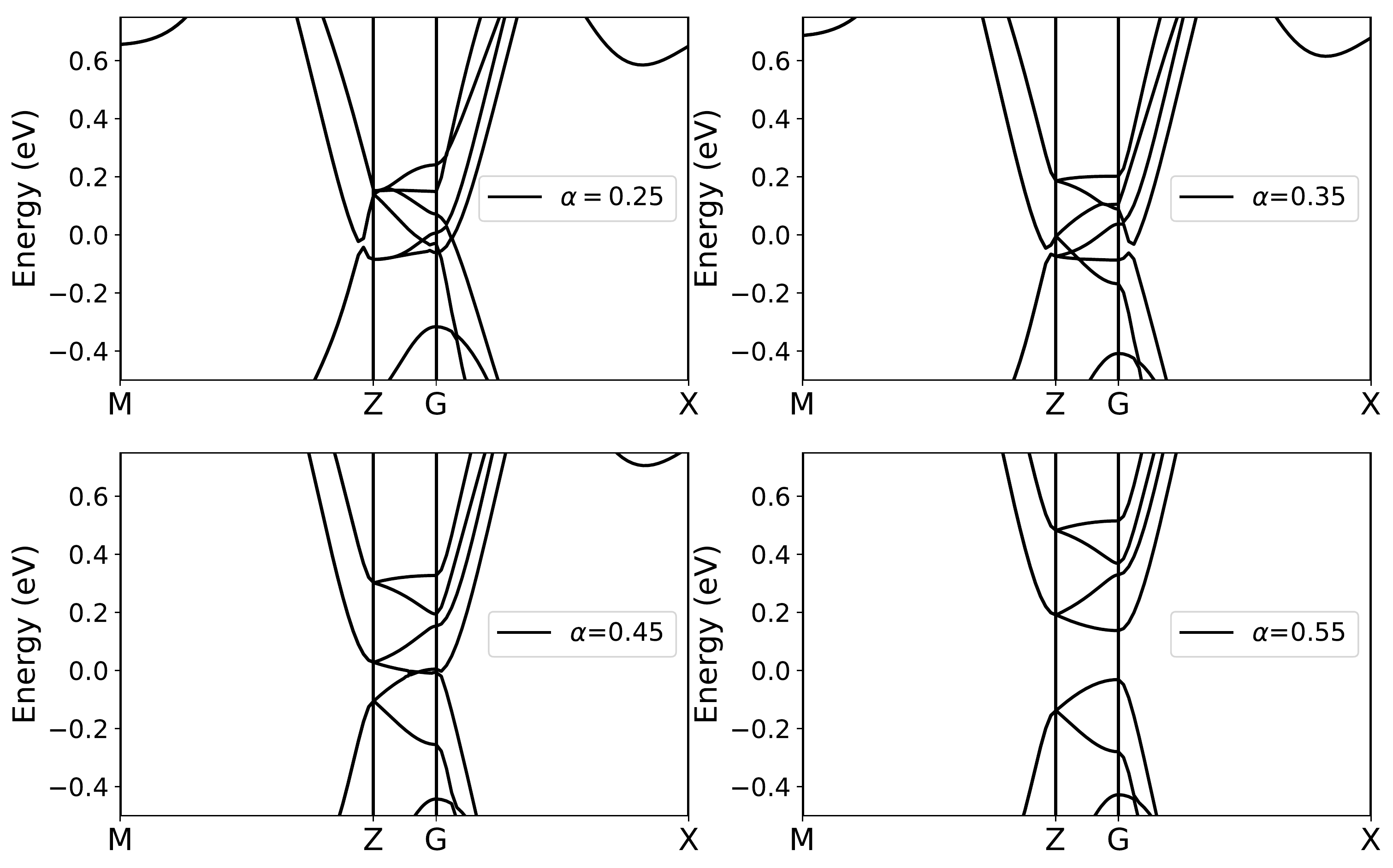}
    \caption{Bandgap opening as we increase the amount of exchange increasing the $\alpha$-parameter in the HSE hybrid functionals in the orthorhombic phase. The gap opens with increasing amount of exchange considered. The HSE03 hybrid functional corresponds to $\alpha=0.25$. In all calculations the range separation parameter is chosen to 0.3}
    \label{fig: alphaEvolution}
\end{figure*}

%\newpage
%\subsection{\label{A: TNSphonons}Phonon spectrum for Ta$_2$NiS$_5$}

\begin{figure*}
    \centering
    \includegraphics[scale=0.40]{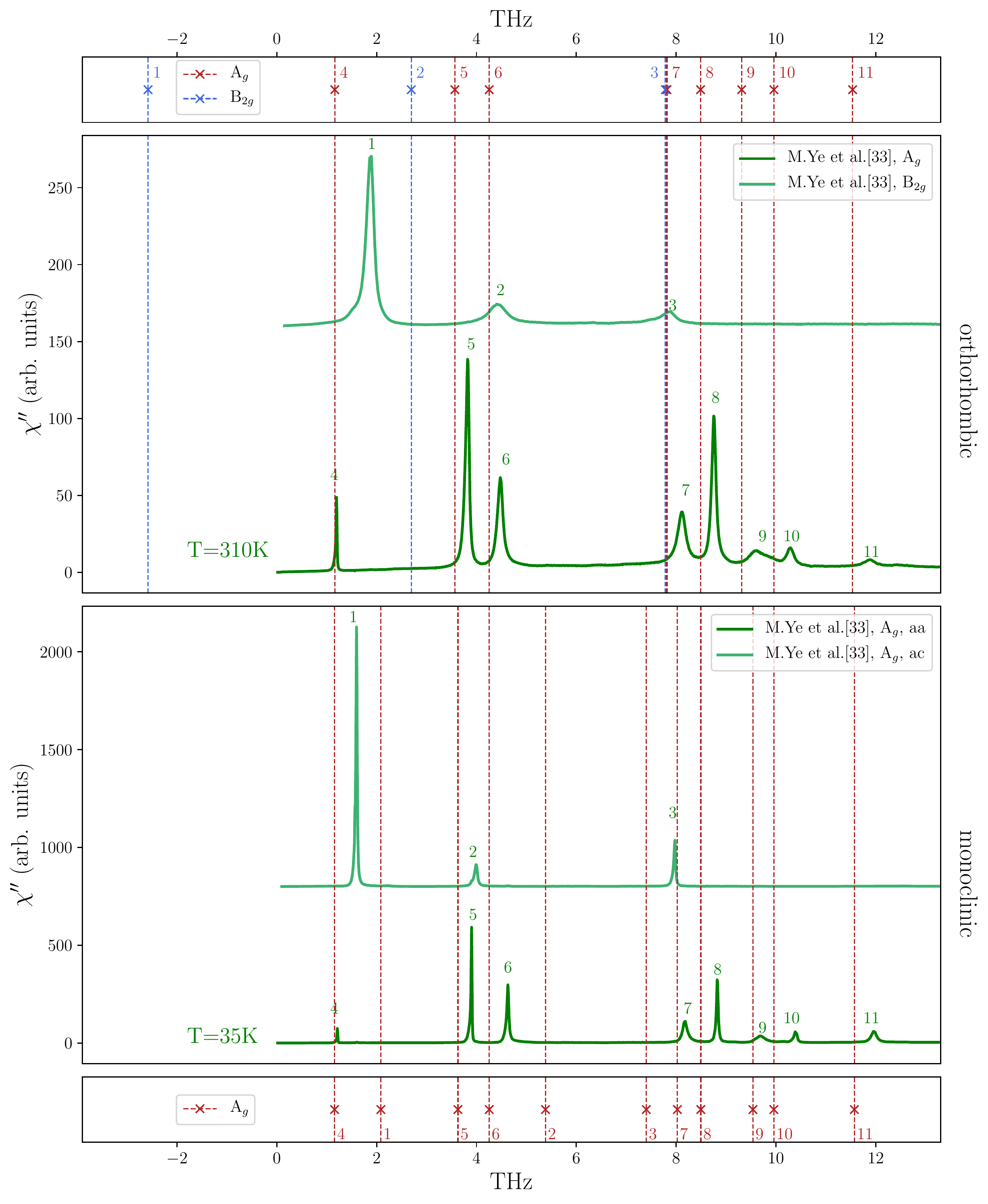}
    \caption{Comparison of the theoretically calculated phonon eigenenergies at T=0K with the Raman spectra provided by M.Ye et al. \cite{Blumberg2021} plotted on a linear scale. The top two panels show the orthorhombic phonon spectra and the bottom two panels show the monoclinic phonon spectra. The theory spectra are obtained at T=0K and are in good agreement with Raman spectra. Only the first two B$_2g$-modes show a relevant discrepancy in their eigenenergy.}
    \label{fig: Raman comparison TNSu}
\end{figure*}

\clearpage
\clearpage

\begin{figure*}%[h!]
    \centering
    \includegraphics[scale=0.35]{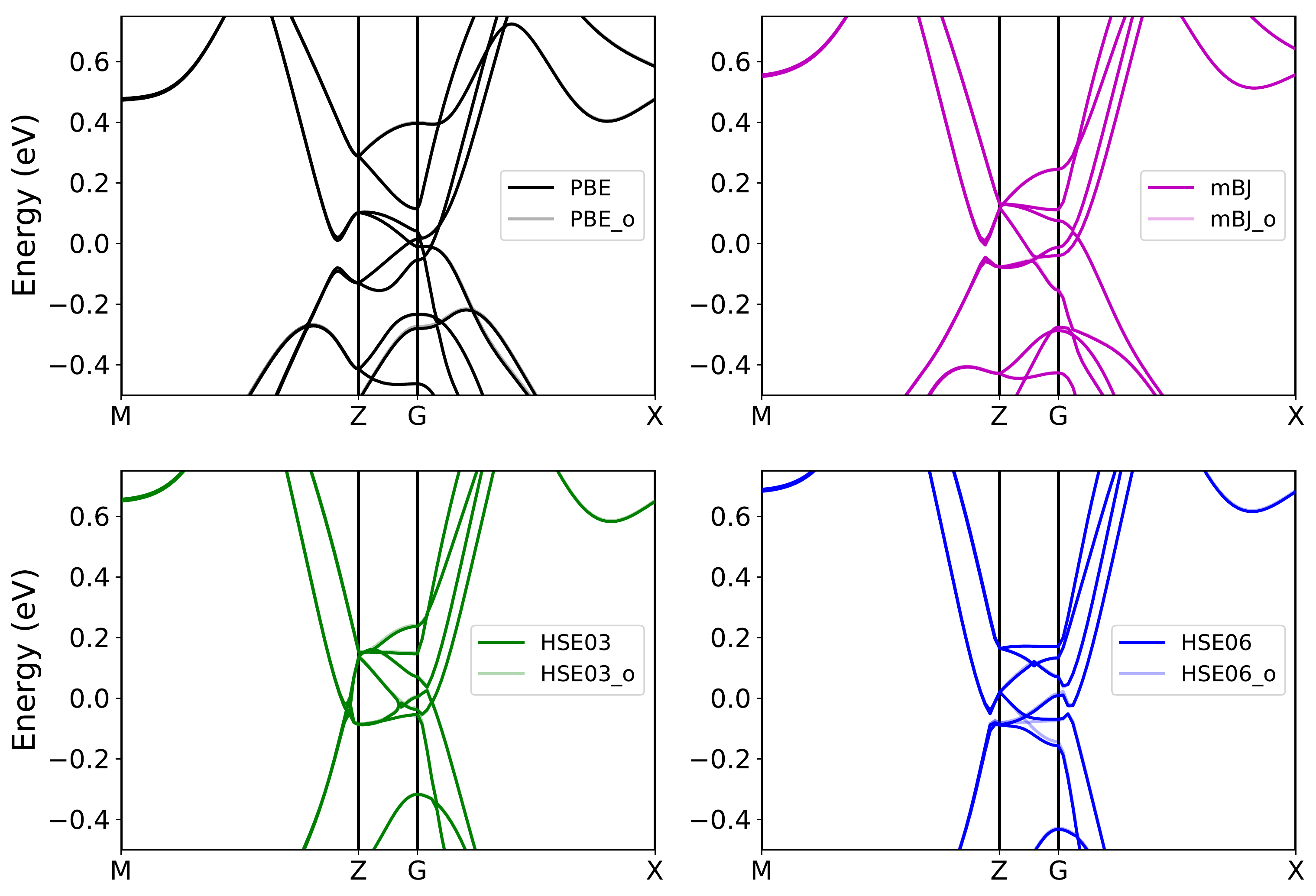}
    \caption{Electronic Dispersion of TNSe for minimally distorted lattice geometry to break the lattice symmetries for different exchange correlation functionals (lattice distortion parameter $d=0.05$). In shallow colors the exact orthorhombic bandstructure is displayed. We see, that for all exchange correlation functionals orthorhombic and symmetry broken dispersions are almost degenerate. We do not observe a metal to semiconductor transition or a significant gap opening for any functional.}
    \label{fig: smallDistortion}
\end{figure*}

%\subsection{\label{A: HSE06}HSE06 calculations}

\begin{figure*}%[h!]
    \centering
    \includegraphics[scale=0.4]{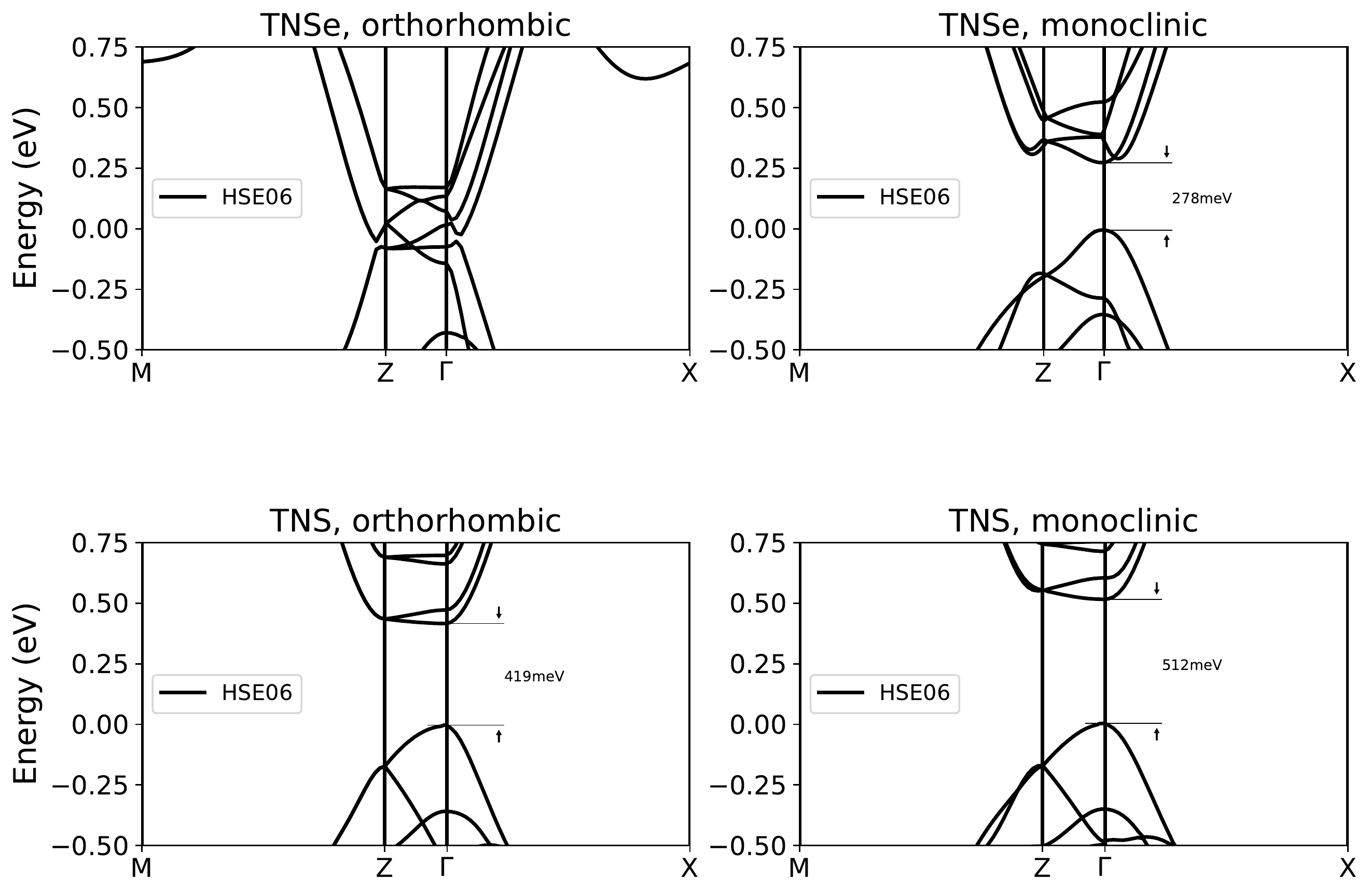}
    \caption{Electronic Dispersion of TNSe and TNS for the orthorhombic and monoclinic geometry using the HSE06 functional.}
    \label{fig: HSE06}
\end{figure*}

%\subsection{G$_0$W$_0$ starting points}
%In this section we present the result of G$_0$W$_0$ calculations starting also from more different hybrid functional calculations (HSE06, PBE0, B3LYP). They all show a similar trend as the HSE03 functional, with the hybrid functional overestimating the bandgap and the following G$_0$W$_0$ correction shifting it to smaller values between 273 meV and 163 meV (see figure \ref{fig: GWstartingPoints}). 

\begin{figure*}%[h!]
    \centering
    \includegraphics[scale=0.4]{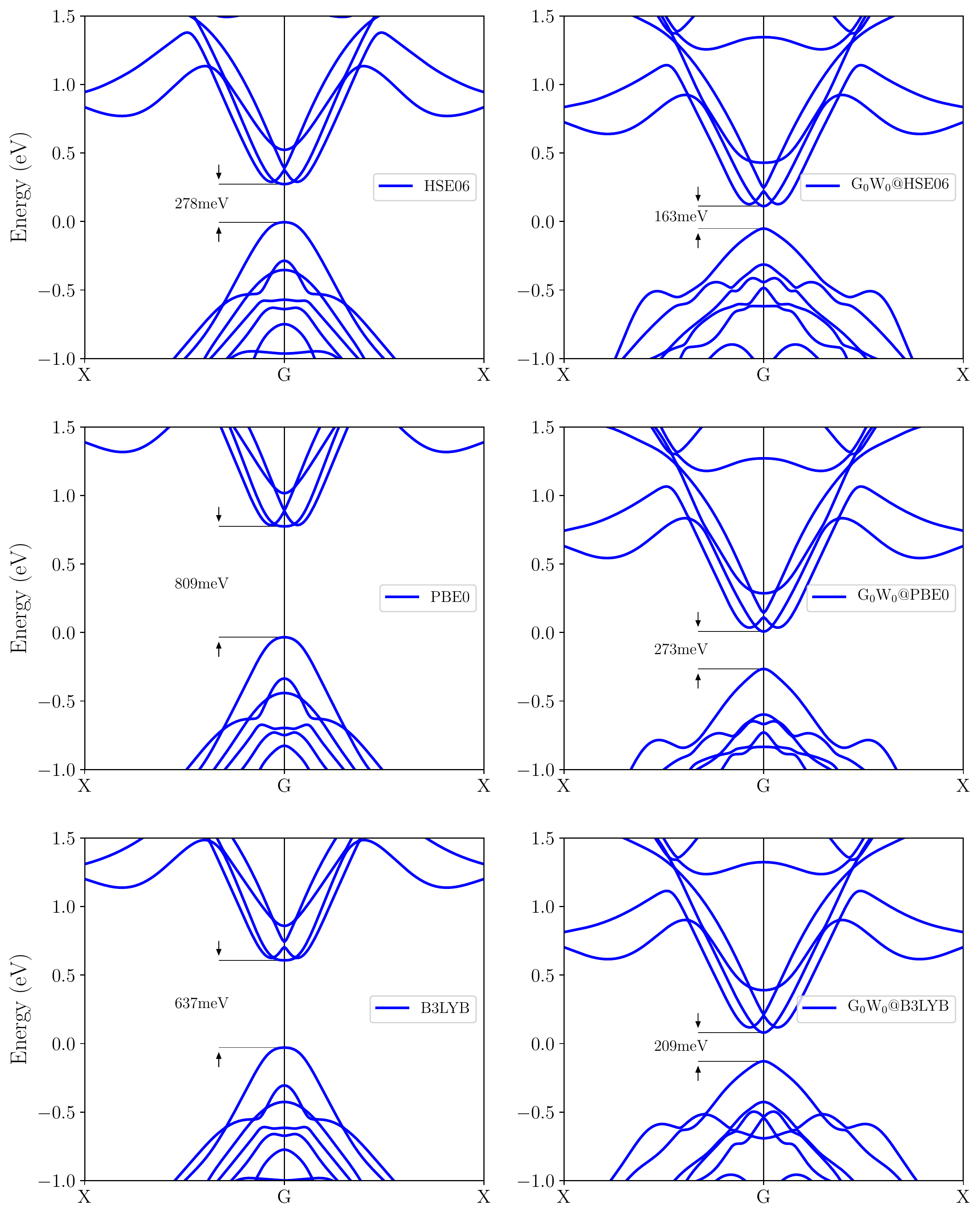}
    \caption{Electronic Dispersion of TNSe for the monoclinic geometry using the G$_0$W$_0$ correction for different hybrid functional starting points.}
    \label{fig: GWstartingPoints}
\end{figure*}

%\subsection{\label{A: HF}Hartree-Fock calculations}
%We have performed Hartree-Fock calculations for both the orthorhombic and monoclinic geometry. The results are displayed in Supplementary Figure~\ref{fig: HF}. The bandgaps are 438 meV for the orthorhombic and 836 meV for the monoclinic geometry.

\begin{figure*}
    \centering
    \includegraphics[scale=0.4]{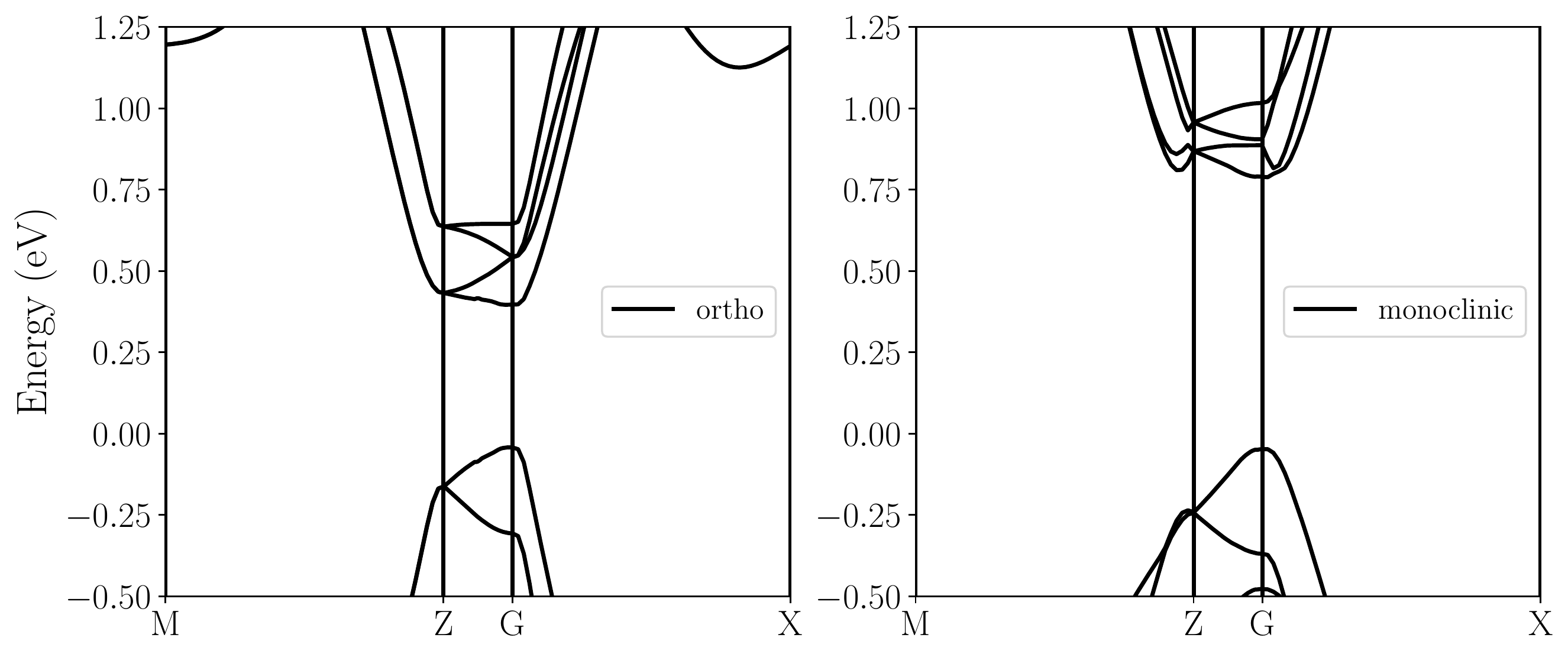}
    \caption{Electronic Dispersion of TNSe for the orthorhombic and monoclinic geometry of TNSe using the Hartree-Fock method. The bandgaps are 438 meV for the orthorhombic and 836 meV for the monoclinic phase. Therefore, Hartree-Fock overestimates the bandgaps significantly.}
    \label{fig: HF}
\end{figure*}

\clearpage

%\subsection{\label{PBE_Froz Phonon}PBE functional, frozen phonon}
%For comparison and completeness we present here the frozen phonon bandstructures of the four Raman active modes presented in section 5 of the main text after DFT calculation using the PBE functional. The result is depicted in Supplementary Figure~\ref{fig: e-p coupling}. We observe a similar behavior as in the G$_0$W$_0$ calculations, but notice a decreased group velocity of the electrons around the bandedge in the plain DFT bandstructures. 

\begin{figure*}%[h!]
    \centering
    \includegraphics[scale=0.45]{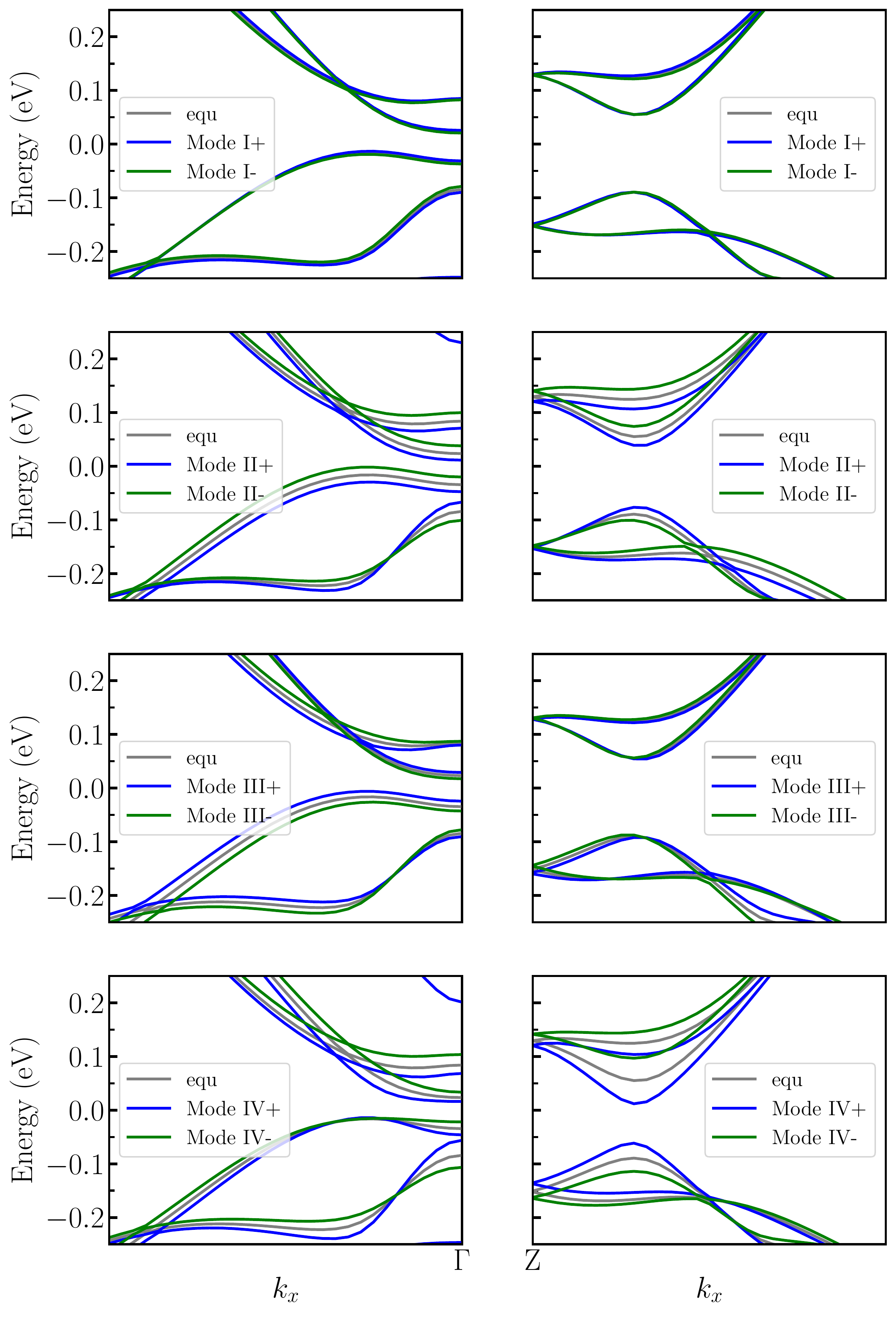}
    \caption{Frozen phonon bandstructures after DFT calculation using the PBE functional. We displayed the result for the four Raman active phonons presented in section 5 of the main text. They show similar behavior as the results after G$_0$W$_0$ calculation. In the G$_0$W$_0$ calculation, however, the group velocity of the electrons is increased.}
    \label{fig: e-p coupling}
\end{figure*}

%\newpage
\clearpage

\begin{figure*}%[h!]
    \centering
    \includegraphics[scale=0.35]{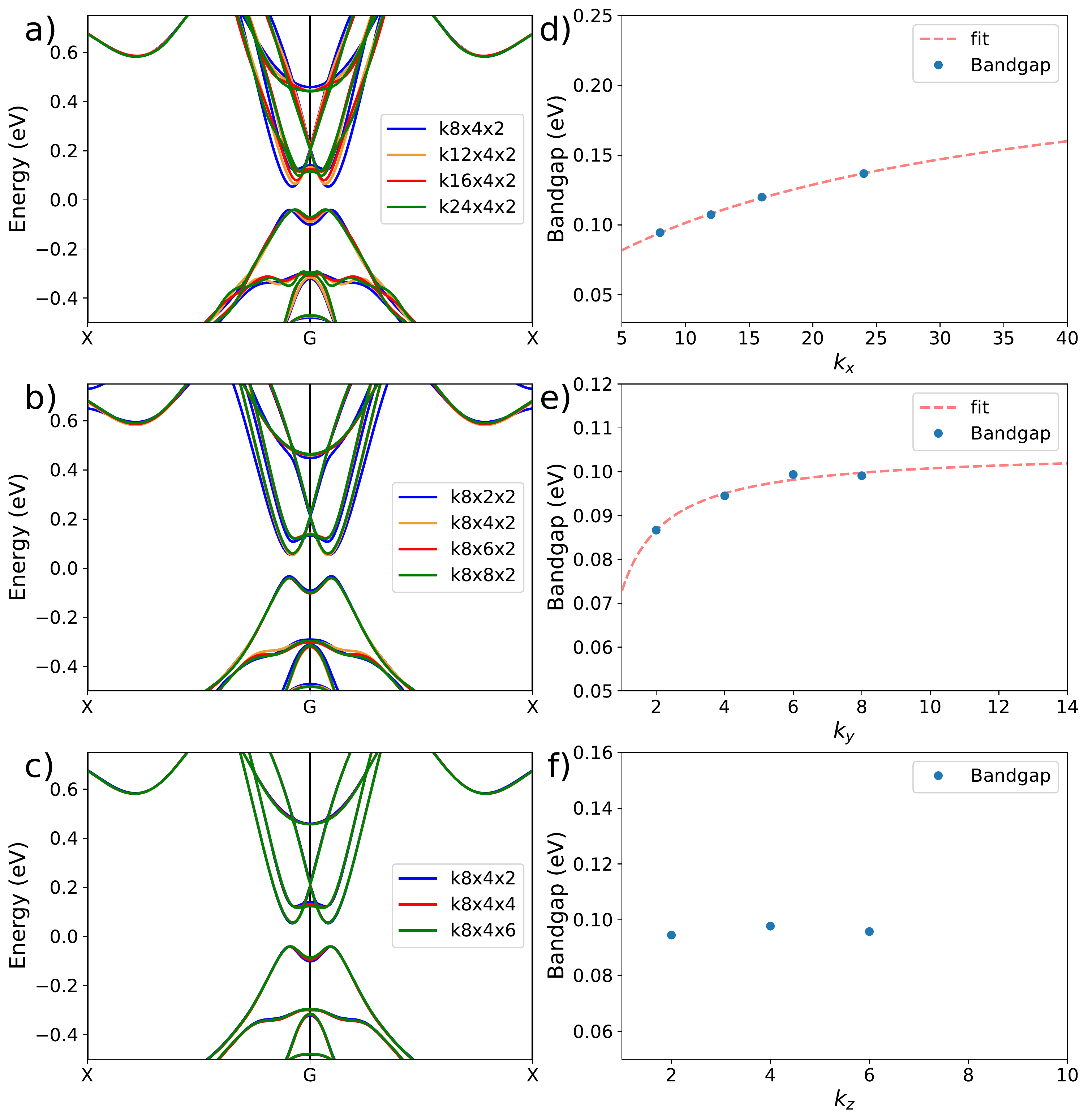}
    \caption{Convergence of the G$_0$W$_0$ bandgap with the k-gridsize. Convergence in each direction is tested separately using 640 unoccupied states, a 80eV Cutoff for the response function and 160 frequency grid points. The panels a) - c) show the corresponding bandstructures computed with an increasing k-mesh into the different directions. the panels d)-f) show the convergence of the bandgap using increasing meshes. $k_x$, $k_y$ an $k_z$ count the number of k-points along the corresponding reciprocal lattice vector. In red dashed lines an extrapolation is displayed. In $k_z$ direction we have too few data points for an extrapolation, but the 3 data points suggest, that convergence is already achieved for 2 points in this direction}
    \label{fig: kConvergence}
\end{figure*}

\begin{figure}%[h!]
    \centering
    \includegraphics[scale=0.43]{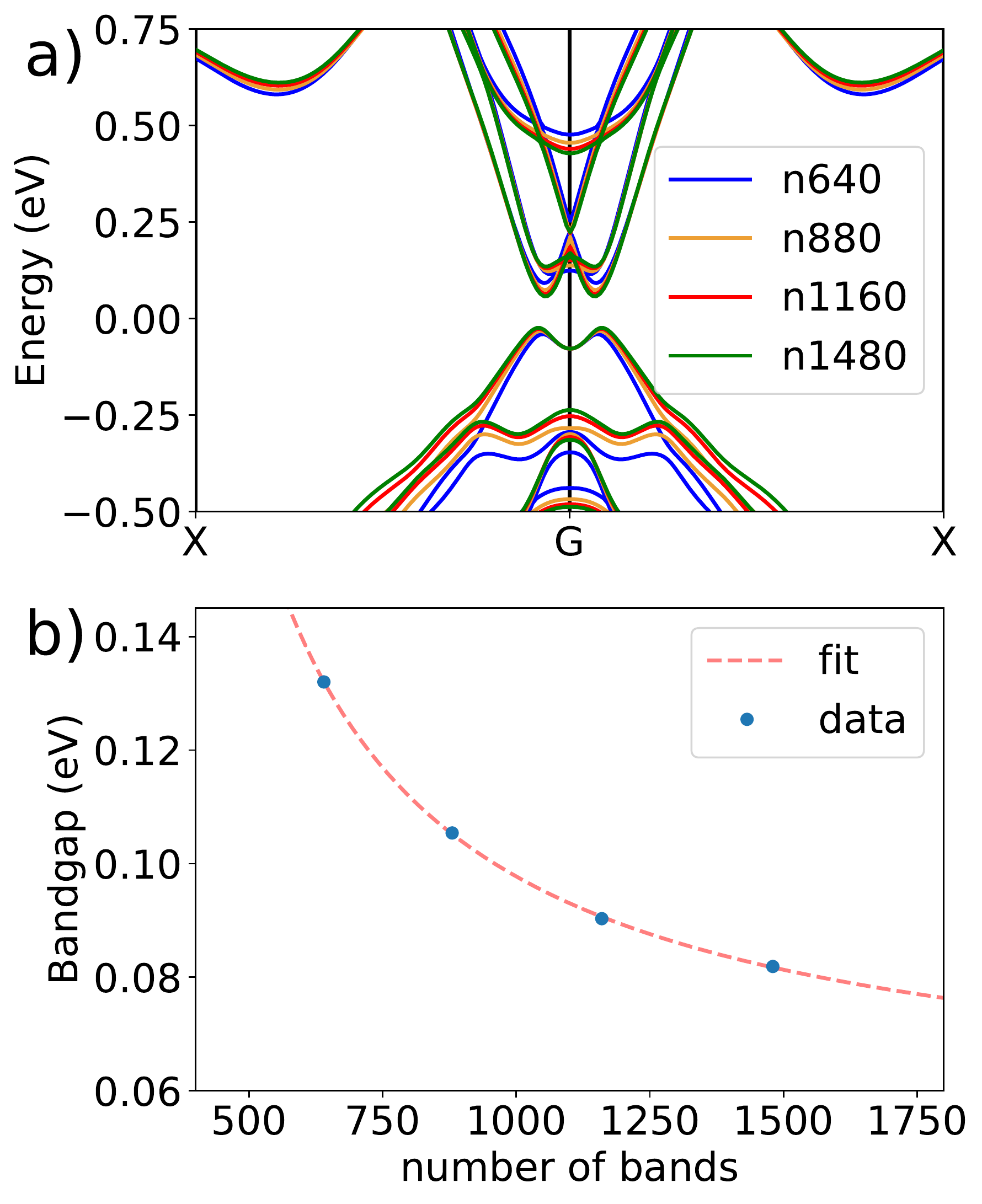}
    \caption{G$_0$W$_0$ convergence using an increasing amount of unoccupied states while also increasing the response function cutoff simultaneously. Panel a) shows the bandstructures for an increasing number of states included in the response function calculation. Panel b) shows the convergence of the bandgap for an increasing number of orbitals included.}
    \label{fig: eCut}
\end{figure}

\begin{figure}%[h!]
    \centering
    \includegraphics[scale=0.26]{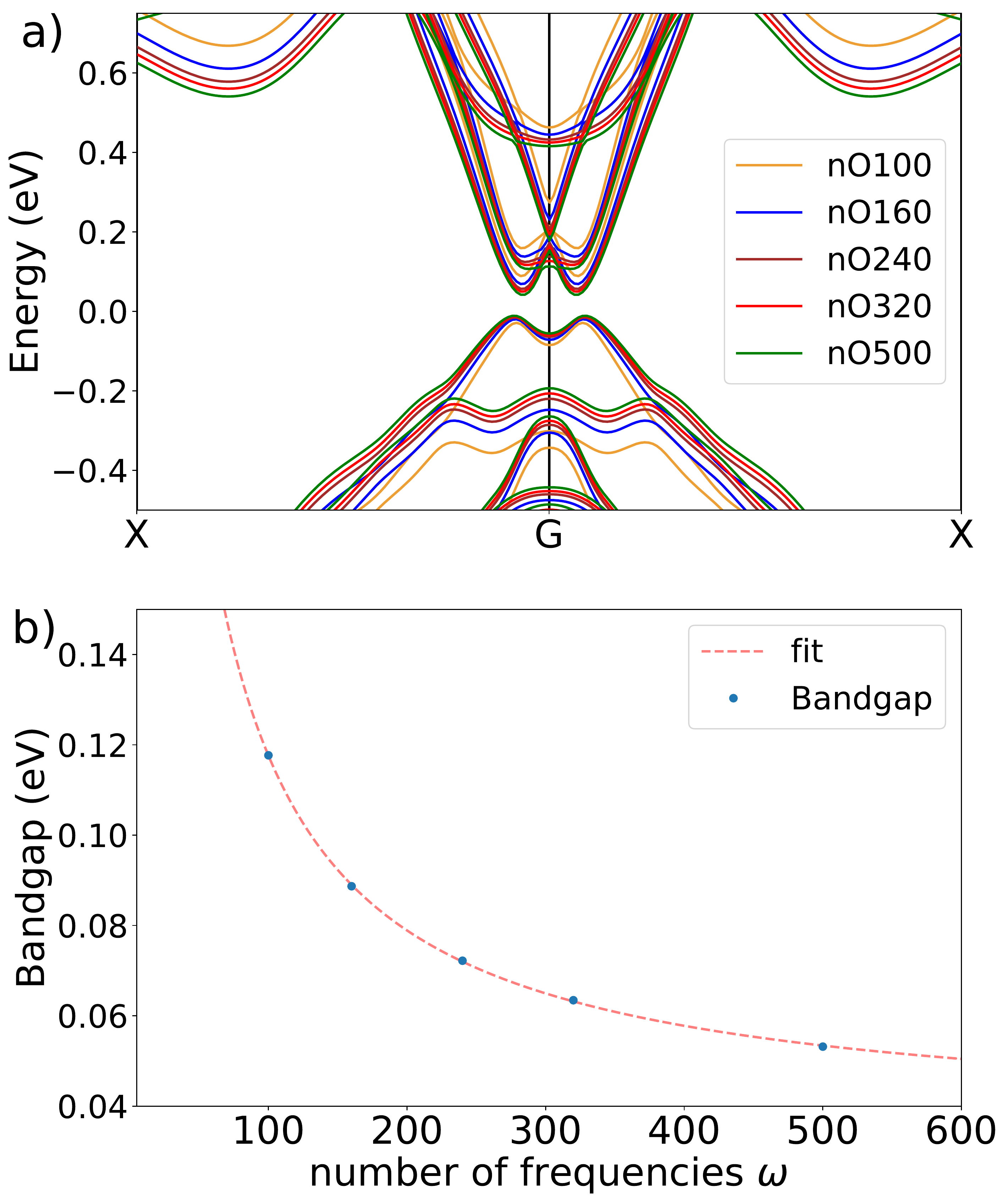}
    \caption{G$_0$W$_0$ convergence using an increasing amount of frequencies $\omega$. Panel a) shows the bandstructures for an increasing number of frequencies included in the screened interaction calculation. Panel b) shows the convergence of the bandgap.}
    \label{fig: nOmega}
\end{figure}

\clearpage
\clearpage

\newpage

\begin{figure*}%[h!]
    \centering
    \includegraphics[scale=0.35]{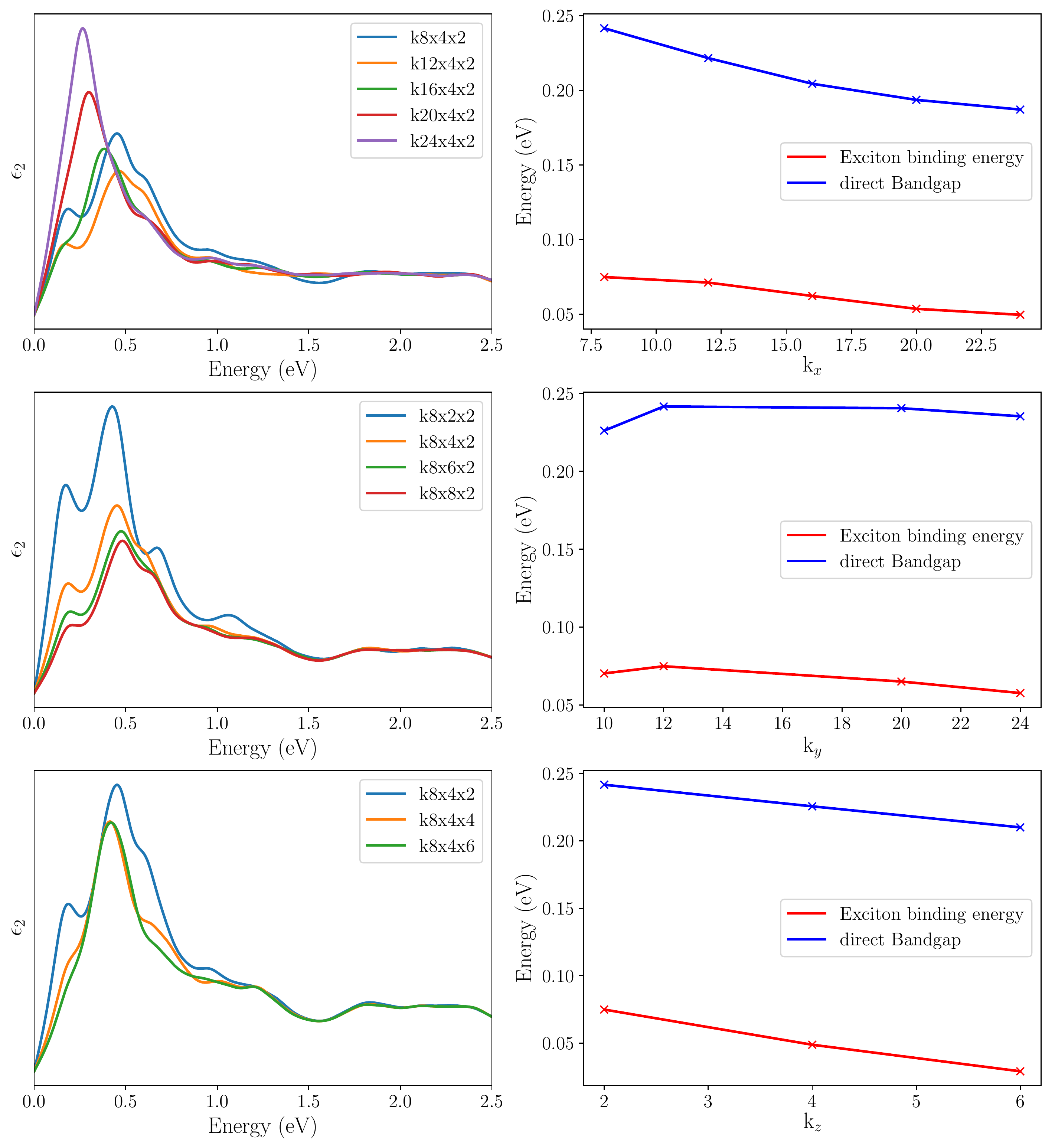}    
    \caption{Convergence of the dielectric function and BSE eigenenergies after solving the BSE using different number of k-points along the 3 axes. We see, that we need many k-Points in x-direction to obtain a converged BSE calculation. This makes sense, as we need a fine grid sampling in this direction to probe the bandgap extrema as well as its highly dispersive character in x-direction.}
    \label{fig: BSE_kConvergence}
\end{figure*}

\begin{figure*}%[h!]
    \centering
    \includegraphics[scale=0.35]{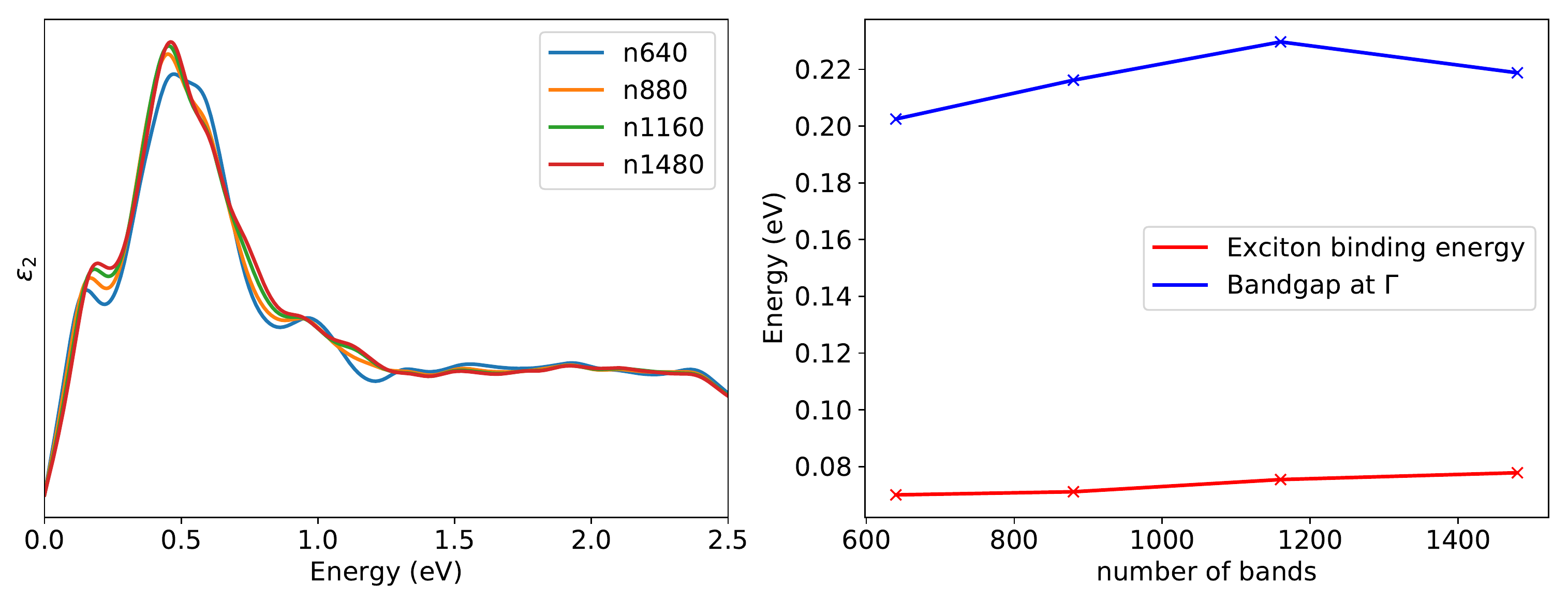} 
    \caption{Convergence of the dielectric function and BSE eigenenergies after solving the BSE using different number of unoccupied states for the calculation of the screened interaction and G$_0$W$_0$ calculation.}
    \label{fig: BSE_CutoffConvergence}
\end{figure*}

\begin{figure*}%[h!]
    \centering
    \includegraphics[scale=0.35]{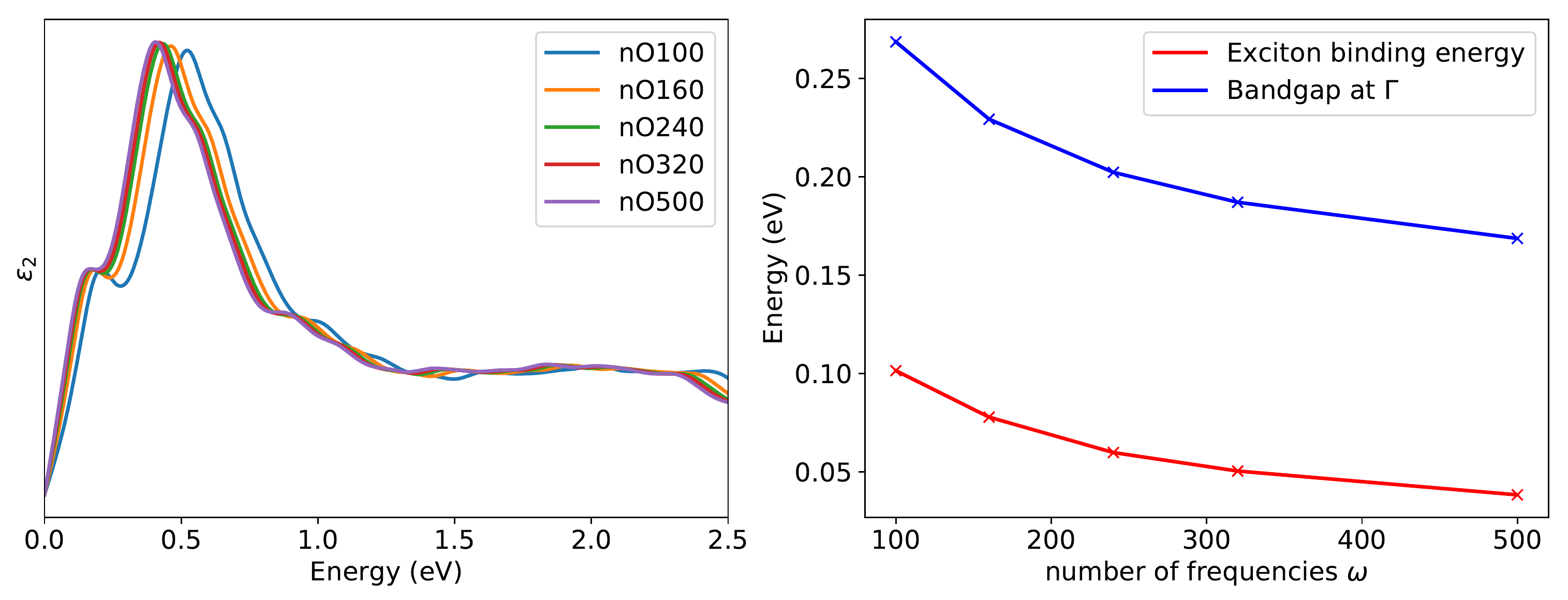}
    \caption{Convergence of the dielectric function and BSE eigenvalues after solving the BSE using different number of frequencies for the computation of the screened interaction and the G$_0$W$_0$ calculation}
    \label{fig: BSE_omegaConvergence}
\end{figure*}

\begin{figure*}%[h!]
    \centering
    \includegraphics[scale=0.35]{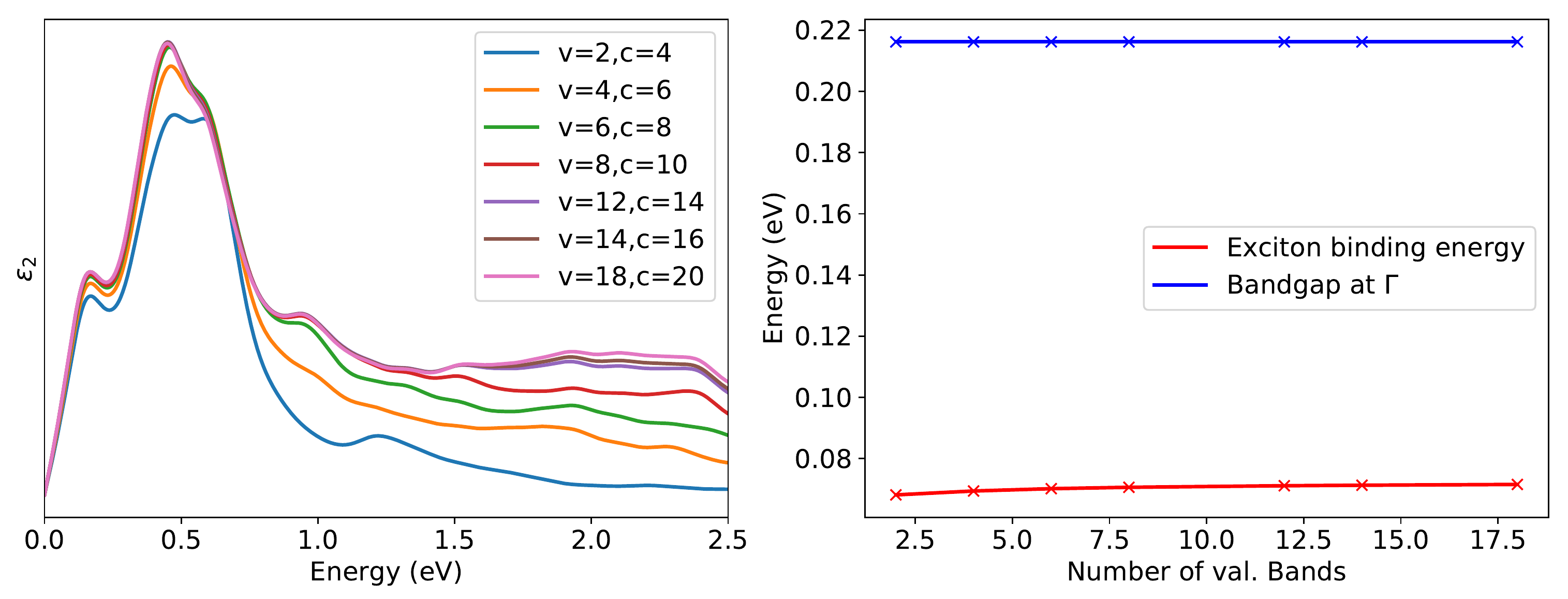}    
    \caption{Convergence of the dielectric function and first exciton energy after solving the Bethe-Salpether equation including an increasing number of valence and conduction bands. We see, that including the first 12 valence and 14 conduction bands results in a well converged dielectric function and excitonic energies.}
    \label{fig: BSE_nVC}
\end{figure*}

\clearpage
\clearpage

\normalem
%bibliographystyle{naturemag}
\bibliography{TNS.bib}

\end{document}